\PassOptionsToPackage{table, xcdraw}{xcolor}

\documentclass[sigconf]{acmart}

\usepackage{color}
\usepackage{soul}    % highlighting
\usepackage{gensymb} % degree symbol

\usepackage{graphicx}

\usepackage{pifont}% http://ctan.org/pkg/pifont
\newcommand{\cmark}{\ding{51}}%
\usepackage{tikz}

\graphicspath{ {figures/} }

\usepackage{enumitem} % remove indent of list items

\usepackage{xspace} % for putting space when necessary

\newcommand{\systemName}{\textsc{TinkerXR}\xspace}

\usepackage{soul}
\AtBeginDocument{%
  }

\copyrightyear{2025}
\acmYear{2025}
\setcopyright{cc}
\setcctype{by-nc}
\acmConference[SCF '25]{ACM Symposium on Computational
Fabrication}{November 20--21, 2025}{Cambridge, MA, USA}
\acmBooktitle{ACM Symposium on Computational Fabrication (SCF '25),
November 20--21, 2025, Cambridge, MA, USA}
\acmDOI{10.1145/3745778.3766651}
\acmISBN{979-8-4007-2034-5/2025/11}

\begin{document}

%%
%% The "title" command has an optional parameter,
%% allowing the author to define a "short title" to be used in page headers.
\title[\systemName: In-Situ, Reality-Aware CAD and 3D Printing Interface for Novices]{\systemName: In-Situ, Reality-Aware CAD and 3D Printing \\ Interface for Novices}

\author{Oğuz Arslan}
\affiliation{%
  \institution{Boğaziçi University}
  \city{Istanbul}
  \country{Türkiye}}
\email{oguz.arslan1@boun.edu.tr}

\author{Artun Akdoğan}
\affiliation{%
  \institution{Boğaziçi University}
  \city{Istanbul}
  \country{Türkiye}}
\email{artun.akdogan@boun.edu.tr}

\author{Mustafa Doga Dogan}
\affiliation{
  \institution{Adobe Research}
  \city{Basel}
  \country{Switzerland}}
\affiliation{%
  \institution{Boğaziçi University}
  \city{Istanbul}
  \country{Türkiye}}
\email{doga@adobe.com}

%%
%% By default, the full list of authors will be used in the page
%% headers. Often, this list is too long, and will overlap
%% other information printed in the page headers. This command allows
%% the author to define a more concise list
%% of authors' names for this purpose.
\renewcommand{\shortauthors}{Arslan et al.}

%%
%% The abstract is a short summary of the work to be presented in the
%% article.
\begin{abstract}

Despite the growing accessibility of augmented reality (AR) for visualization, existing computer-aided design (CAD) systems remain confined to traditional screens or require complex setups or predefined parameters, limiting immersion and accessibility for novices. We present \systemName, an open-source AR interface enabling in-situ design and fabrication through Constructive Solid Geometry (CSG) modeling. \systemName operates solely with a headset and 3D printer, allowing users to design directly in and for their physical environments. By leveraging spatial awareness, depth occlusion, recognition of physical constraints, reference objects, and hand movement controls, \systemName enhances realism, precision, and ease of use. Its AR-based workflow integrates design and 3D printing with a drag-and-drop interface for printers' virtual twins.

A user study comparing \systemName with Tinkercad shows that \systemName offers novices higher accessibility, engagement, and ease of use. Participants highlighted how designing directly in physical space made the process more intuitive.
By bridging the gap between digital creation and physical output, \systemName aims to transform everyday spaces into expressive creative studios. We release \systemName as open source\footnote{\systemName is available at \url{http://tinkerxr.github.io}.} to encourage further exploration of accessible, spatially grounded CAD tools.

\end{abstract}

%%
%% The code below is generated by the tool at http://dl.acm.org/ccs.cfm.
%% Please copy and paste the code instead of the example below.
%%
%%


%%
%% Keywords. The author(s) should pick words that accurately describe
%% the work being presented. Separate the keywords with commas.
%% A "teaser" image appears between the author and affiliation
%% information and the body of the document, and typically spans the
%% page.
\begin{teaserfigure}
  \includegraphics[width=1\textwidth]{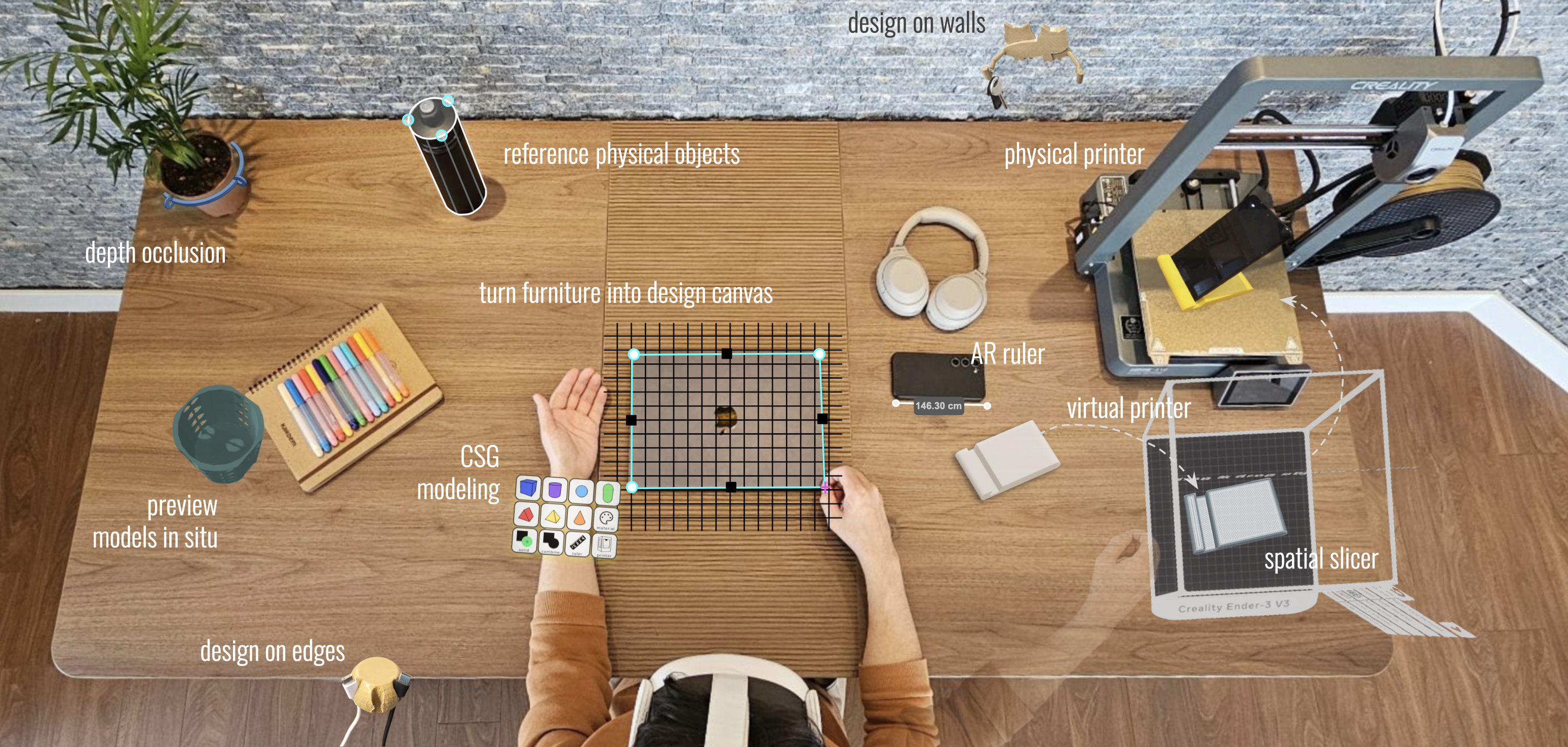}
  \caption{\systemName is an in-situ CAD and 3D printing AR interface. The figure illustrates key features such as spatial awareness, depth occlusion, and the use of physical objects and constraints for design. With \textit{Constructive Solid Geometry} modeling, users can convert surfaces into interactive design canvases, %employ AR rulers for precision,
  and after designing, interact with 3D printers' digital twins to prepare for and initiate fabrication. Examples of designs showcased in this figure include a pen holder, phone stand, cable organizer, plant pot holder, bottle mount, and key hook.}
  \Description{Five images labeled (a), (b), (c), (d), and (e) show the user interacting with the system. In pictures (a), (b), and (c), the user is seen wearing a headset. In picture a, the user is in a room with a desk. The user is standing up next to the desk and pointing their finger at the desk. A ray from their finger hits the table, where a text appears on top, saying "select: table". Three similar lines are extended towards the floor, wall, and ceiling. In pictures (b) and (c), the user is sitting on the desk, where a grid is seen covering the top surface of the desk. In the picture (b), the user is creating a cube object with a poking gesture with their right hand on the menu on the inside of their left wrist. In picture (c), the user is holding a real-world finger skateboard on their left hand while applying a pinch gesture with a right hand onto a virtual finger skateboard obstacle. In picture (d), the first-person view of the user is seen where a virtual 3D printer twin is on top of the desk and the user is dragging the designed model with their left hand on the printer twin. The user is poking the button next to the digital printer twin labeled "Start print". In picture (e), the 3D-printed result can be seen on top of the desk.}
  \label{fig:teaser}
\end{teaserfigure}

%%
%% This command processes the author and affiliation and title
%% information and builds the first part of the formatted document.
\maketitle

\section{Introduction}
Augmented reality (AR) has emerged as a transformative technology across various fields, notably in design and manufacturing. Traditional computer-aided design (CAD) tools such as \textit{SolidWorks} and Blender have provided high levels of precision and functionality \cite{reyes_beginners_guide_to_solidworks_2018, mullen_mastering_2012}, while platforms like \textit{Tinkercad} have made CAD more accessible to novices and educational users \cite{jacob_use_2021, mahapatra_barriers_2019}. Despite the advances in AR, however, these tools are still mostly limited to traditional 2D screens, which limits their interactivity and integration with the user's real-world environment.

The introduction of \textit{Tinkercad AR}\footnote{\url{https://www.tinkercad.com/ipad-app}, Accessed: 16.07.2024} marked a notable progression by incorporating AR to view designs in physical spaces, albeit mainly for visualization. Applications such as \textit{Gravity Sketch}\footnote{\url{https://www.gravitysketch.com/}, Accessed: 16.07.2024} have begun to leverage virtual reality (VR) headsets and controllers to enable more immersive 3D sketching experiences. In recent HCI research, several notable advancements have emerged to enhance spatial design interfaces. \textit{pARam} \cite{stemasov_param_2024} suggests that in-situ parametric design and customization can be effectively achieved through hand interactions and environmental considerations in AR. Additionally, \textit{BrickStARt} \cite{stemasov_brickstart_2023} introduces a novel approach by integrating tangible construction blocks with mixed reality (MR) to enable users to design and test physical artifacts directly in their intended locations. \textit{ProtoAR} \cite{nebeling_protoar_2018} addresses the challenges of AR content creation by combining physical prototyping with mobile authoring tools for rapid development and interaction. Furthermore, \textit{MixFab} \cite{weichel_mixfab_2014} illustrates how MR can streamline personal fabrication by facilitating direct interaction with virtual objects within an AR environment. Collectively, these advancements highlight a trend towards more spatially aware design interfaces, which are particularly beneficial for complex tasks that integrate digital and physical spaces~\cite{ProgrammableReality25}.

Despite these advancements, a significant gap remains in harnessing AR for interactive design that directly integrates with the user’s physical environment and supports seamless transitions from modeling in AR to 3D printing.
Focusing on this gap, we introduce \systemName, an AR interface that enables \textbf{in-situ CAD and 3D printing for novices}. Our system bridges the gap between simple, novice-friendly tools and immersive, spatial capabilities of advanced AR technologies, offering spatial interaction techniques for designing 3D models directly in the physical environment using only hand controls. \systemName employs \textit{Constructive Solid Geometry} modeling with primitives similar to those in \textit{Tinkercad}, but enhances the design experience by integrating these elements within the user’s physical environment, integrating a 3D printing workflow, and making the design process more adaptable to the real world through AR.

\begin{figure*}[h]
  \centering
  \includegraphics[width=1\textwidth]{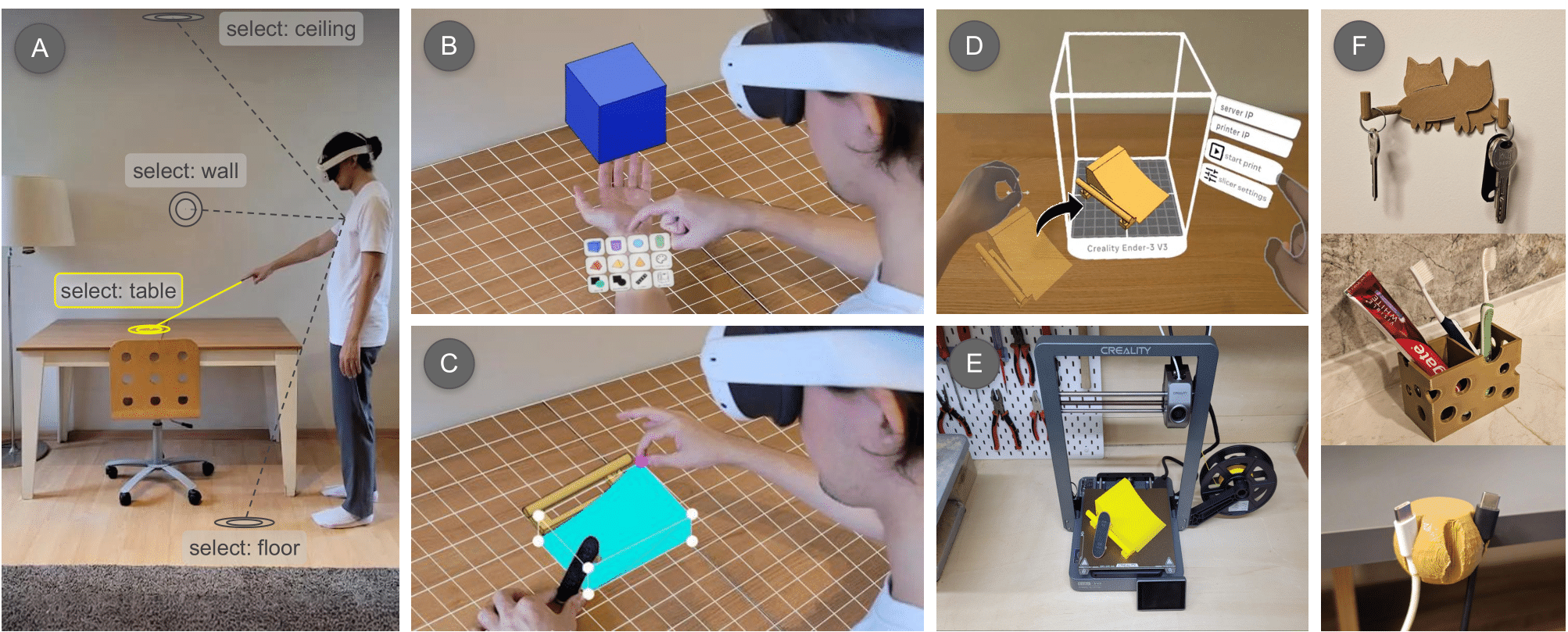}
  \caption{\textbf{User workflow}. (a) Users begin by selecting a workspace from one of the flat surfaces in their room. (b, c) On the selected surface, users can create and manipulate primitive geometries using their hands while utilizing the tools provided by \systemName. In this example, the user designs a finger skateboard obstacle. (d) The final model is then placed in a virtual 3D printer twin where print slicing options can be adjusted before (e) fabricating on the remote physical printer. (f) Examples from the user study (Section~\ref{user-study}) that demonstrate the \textit{design anywhere} functionality, i.e., the flexibility to adapt to different environmental contexts.
  }
  \Description{}
  \label{fig:implementation_teaser}
\end{figure*}

Our system utilizes a state-of-the-art headset\footnote{\textit{Meta Quest 3}: \url{https://www.meta.com/quest/quest-3/}, Accessed: 16.07.2025}  and hand-gesture controls as well as passthrough features, enabling users to \textit{interact with and create 3D models \textbf{directly atop} their actual desks or on other flat surfaces like walls or the ground} (\autoref{fig:implementation_teaser}a). This setup facilitates interaction with the digital content, and also allows users to accurately gauge the size and scale of their designs in relation to their immediate surroundings.

Building upon prior work in environmentally-aware AR interfaces \cite{stemasov_param_2024, stemasov_brickstart_2023, weichel_mixfab_2014}, \systemName is designed to leverage \textbf{scene understanding} to turn everyday surfaces into functional CAD workspaces. By incorporating spatial features such as \textbf{snapping to environmental grids} and \textbf{referencing real-world objects}, our system supports a spatially-aligned modeling workflow. To create a more coherent and visually realistic MR experience, it uses depth-sensing to ensure virtual objects are accurately occluded by real-world elements.

Furthermore, \systemName extends the concept of AR in CAD beyond digital output to include actual manufacturing processes. It integrates directly with 3D printing by incorporating a \textbf{\textit{drag-and-drop metaphor}} that allows users to quickly prepare models for fabrication. 
%doga{does this sentence make sense? ->}
Acting as a built-in slicing interface in AR, \systemName prepares models for 3D printing while providing real-time feedback through a virtual 3D printer twin, which allows users to make adjustments and visualize the output in real-time.

To evaluate \systemName, we conducted a user study (\textit{n}=10) comparing it against \textit{Tinkercad} as a baseline for comparison. The participants completed three design tasks and a 3D print preparation task in both systems. We found no statistically significant difference in design completion time, while our integrated 3D printing workflow made the \textbf{print preparation process faster} than the traditional export-and-slice method. Subjectively, participants rated \systemName as having \textbf{higher usability} (SUS) \cite{brooke_sus_1996} and reported a \textbf{lower mental workload} (NASA-TLX) \cite{hart_development_1988}, with the primary trade-off being a higher physical workload, which we discuss in the results.

\vspace{0.15cm}

Overall, we present a novel design and fabrication system that integrates  in-situ features with hand control schemes and a novice-friendly design logic. By bridging the digital and physical worlds, our system aims to enable seamless design-to-manufacturing workflows. We summarize our contributions as follows:

\vspace{-0.03cm}

\begin{itemize}[leftmargin=0.4cm]
    \item We present \systemName, an in-situ CAD system that enables hand-based \textbf{CSG modeling} directly in \textit{any} physical environment, tailored for novices. The system's novel contributions include interaction methods specifically designed for CSG design in real-world settings, such as snap-to-grid, designing on top of reference objects, as well as novice-friendly controls for applying traditional CAD functionalities with hand interactions.
    \item We integrate \textbf{3D printing workflows} directly into the AR design process, allowing users to seamlessly transition from digital design to physical creation. The system includes a built-in \textbf{slicer}, real-time feedback via a \textbf{virtual 3D printer twin}, and the ability to initiate printing directly from the immersive interface.
    \item We present the results of a \textbf{comparative user study} that characterizes the trade-offs between \systemName and \textit{Tinkercad} for novice users. Our findings show that \systemName achieves \textbf{comparable design efficiency, streamlines the fabrication workflow, and reduces mental workload}, at the cost of increased physical demand.
    \item We make \systemName \textbf{open source}\footnote{\systemName is available at \url{http://tinkerxr.github.io}.}, contributing to the research community by providing the system and codebase for further development, experimentation, and possible integration with further fabrication technologies.
\end{itemize}

Our approach simplifies AR-based geometric modeling and supports a flexible methodology for designing in any place, smoothly transitioning from digital designs to physical outputs. By enhancing 3D creativity through a context-aware immersive platform, we hope to advance the field of design and fabrication through AR.

% \DeclareRobustCommand{\hlcyan}[1]{{\sethlcolor{cyan}\hl{#1}}}
% \DeclareRobustCommand{\hlred}[1]{{\sethlcolor{red}\hl{#1}}}
% \DeclareRobustCommand{\hlgreen}[1]{{\sethlcolor{green}\hl{#1}}}

\section{Related Work}

This section reviews the literature in three main areas: Extended reality (XR) in digital design and fabrication, in-situ tools and interfaces, and systems tailored for novice users. We provide a comparative overview of existing AR-based CAD systems, highlighting our system's unique capabilities as summarized in \autoref{tab:relatedWorkComparison}.

% Please add the following required packages to your document preamble:
% \usepackage{graphicx}
% \usepackage[table,xcdraw]{xcolor}
% Beamer presentation requires \usepackage{colortbl} instead of \usepackage[table,xcdraw]{xcolor}
\begin{table*}[]
\centering
\caption{Comparison of features of AR-based CAD interfaces reported in the literature.}
\vspace{-0.25cm}
\label{tab:relatedWorkComparison}
\resizebox{\textwidth}{!}{%
\begin{tabular}{lllllll
>{\columncolor[HTML]{CCE3C3}}l }

& \cellcolor[HTML]{EFEFEF}\textbf{Situated Modeling} 
    & \cellcolor[HTML]{EFEFEF}\textbf{MixFab} 
    & \cellcolor[HTML]{EFEFEF}\textbf{RoMA} 
    & \cellcolor[HTML]{EFEFEF}\textbf{DesignAR} 
    & \cellcolor[HTML]{EFEFEF}\textbf{pARam} 
    & \cellcolor[HTML]{EFEFEF}\textbf{WindowShaping} 
    & \cellcolor[HTML]{EFEFEF}\textbf{TinkerXR} \\
    & \cellcolor[HTML]{EFEFEF}\cite{lau_situated_2012}
    & \cellcolor[HTML]{EFEFEF}\cite{weichel_mixfab_2014}
    & \cellcolor[HTML]{EFEFEF}\cite{peng_roma_2018}
    & \cellcolor[HTML]{EFEFEF}\cite{reipschlager_designar_2019}
    & \cellcolor[HTML]{EFEFEF}\cite{stemasov_param_2024}
    & \cellcolor[HTML]{EFEFEF}\cite{huo_window-shaping_2017}
    &  \cellcolor[HTML]{EFEFEF}[our system] % No reference given for TinkerXR
                                                                                                             % & \cellcolor[HTML]{EFEFEF}\textbf{Situated Modeling %~\cite{lau_situated_2012}
                                                                                                             % }                                           & \cellcolor[HTML]{EFEFEF}\textbf{MixFab  %~\cite{weichel_mixfab_2014}
                                                                                                             % }                                       & \cellcolor[HTML]{EFEFEF}\textbf{RoMA
                                                                                                             % % ~\cite{peng_roma_2018}
                                                                                                             % }                                                                                                  & \cellcolor[HTML]{EFEFEF}\textbf{DesignAR %~\cite{reipschlager_designar_2019}
                                                                                                             % }                                            & \cellcolor[HTML]{EFEFEF}\textbf{pARam %~\cite{stemasov_param_2024}
                                                                                                             % }                                           & \cellcolor[HTML]{EFEFEF}\textbf{WindowShaping %~\cite{huo_window-shaping_2017}
                                                                                                             % }                                    & \cellcolor[HTML]{EFEFEF}\textbf{TinkerXR} 
                                                                                                             \\ \hline
\multicolumn{1}{l|}{\cellcolor[HTML]{f4f4f4}\begin{tabular}[c]{@{}l@{}}\textbf{end-to-end}\\ \textbf{in-situ} CAD\end{tabular}}                        & \multicolumn{1}{l|}{\cellcolor[HTML]{CCE3C3}{\color[HTML]{333333} \cmark}}                      & \multicolumn{1}{l|}{\cellcolor[HTML]{CCE3C3}{\color[HTML]{333333} \cmark}}       & \multicolumn{1}{l|}{\cellcolor[HTML]{CCE3C3}{\color[HTML]{333333} \cmark}}                                                                & \multicolumn{1}{l|}{\cellcolor[HTML]{CCE3C3}{\color[HTML]{333333} \cmark}}              & \multicolumn{1}{l|}{\cellcolor[HTML]{CCE3C3}{\color[HTML]{333333} \cmark}}          & \multicolumn{1}{l|}{\cellcolor[HTML]{CCE3C3}{\color[HTML]{333333} \cmark}}           & {\color[HTML]{333333} \cmark}                \\ \hline
\multicolumn{1}{l|}{\cellcolor[HTML]{f4f4f4}\begin{tabular}[c]{@{}l@{}}built-in \textbf{3D printing}\\ or \textbf{fabrication}\\ workflow\end{tabular}} & \multicolumn{1}{l|}{\begin{tabular}[c]{@{}l@{}}CAD only\\  \end{tabular}}                      & \multicolumn{1}{l|}{\begin{tabular}[c]{@{}l@{}}CAD only \\\end{tabular}}       & \multicolumn{1}{l|}{\cellcolor[HTML]{CCE3C3}{\color[HTML]{333333} \begin{tabular}[c]{@{}l@{}}\cmark\\ wireframe\\ printing\end{tabular}}} & \multicolumn{1}{l|}{\begin{tabular}[c]{@{}l@{}}CAD only \\ \end{tabular}}              & \multicolumn{1}{l|}{\begin{tabular}[c]{@{}l@{}}CAD only\\ \end{tabular}}          & \multicolumn{1}{l|}{\begin{tabular}[c]{@{}l@{}}CAD only\\  \end{tabular}}           & {\color[HTML]{333333} \cmark}                \\ \hline
\multicolumn{1}{l|}{\cellcolor[HTML]{f4f4f4}\begin{tabular}[c]{@{}l@{}}\textbf{CSG}\\ modeling\end{tabular}}                                  & \multicolumn{1}{l|}{\begin{tabular}[c]{@{}l@{}}custom\\ stamping scheme\end{tabular}}        & \multicolumn{1}{l|}{\cellcolor[HTML]{CCE3C3}{\color[HTML]{333333} \cmark}}       & \multicolumn{1}{l|}{\begin{tabular}[c]{@{}l@{}}sketching\\ + extrusion\end{tabular}}                                                   & \multicolumn{1}{l|}{\begin{tabular}[c]{@{}l@{}}sketching\\ + extrusion\end{tabular}} & \multicolumn{1}{l|}{\begin{tabular}[c]{@{}l@{}}parametric\\ design\end{tabular}} & \multicolumn{1}{l|}{\begin{tabular}[c]{@{}l@{}}sketch\\ + inflation\end{tabular}} & {\color[HTML]{333333} \cmark}                \\ \hline
\multicolumn{1}{l|}{\cellcolor[HTML]{f4f4f4}\begin{tabular}[c]{@{}l@{}}design\\ \textbf{anywhere}\end{tabular}}                               & \multicolumn{1}{l|}{\cellcolor[HTML]{CCE3C3}{\color[HTML]{333333} \cmark}}                      & \multicolumn{1}{l|}{\begin{tabular}[c]{@{}l@{}}fixed\\ tabletop\end{tabular}} & \multicolumn{1}{l|}{\begin{tabular}[c]{@{}l@{}}robotic\\ setup\end{tabular}}                                                           & \multicolumn{1}{l|}{\begin{tabular}[c]{@{}l@{}}fixed\\ tabletop\end{tabular}}        & \multicolumn{1}{l|}{\cellcolor[HTML]{CCE3C3}{\color[HTML]{333333} \cmark}}          & \multicolumn{1}{l|}{\cellcolor[HTML]{CCE3C3}{\color[HTML]{333333} \cmark}}           & {\color[HTML]{333333} \cmark}                \\ \hline
\multicolumn{1}{l|}{\cellcolor[HTML]{f4f4f4}\begin{tabular}[c]{@{}l@{}}reference\\ \textbf{physical} objects\end{tabular}}                    & \multicolumn{1}{l|}{\cellcolor[HTML]{CCE3C3}{\color[HTML]{333333} \cmark}}                      & \multicolumn{1}{l|}{\cellcolor[HTML]{CCE3C3}{\color[HTML]{333333} \cmark}}       & \multicolumn{1}{l|}{\begin{tabular}[c]{@{}l@{}}not\\ applicable\end{tabular}}                                                          & \multicolumn{1}{l|}{\cellcolor[HTML]{CCE3C3}{\color[HTML]{333333} \cmark}}              & \multicolumn{1}{l|}{\begin{tabular}[c]{@{}l@{}}not\\ applicable\end{tabular}}    & \multicolumn{1}{l|}{\cellcolor[HTML]{CCE3C3}{\color[HTML]{333333} \cmark}}           & {\color[HTML]{333333} \cmark}                \\ \hline
\multicolumn{1}{l|}{\cellcolor[HTML]{f4f4f4}\begin{tabular}[c]{@{}l@{}}\textbf{depth}\\ occlusion\end{tabular}}                               & \multicolumn{1}{l|}{\begin{tabular}[c]{@{}l@{}}not\\ applicable\end{tabular}}                & \multicolumn{1}{l|}{\cellcolor[HTML]{CCE3C3}{\color[HTML]{333333} \cmark}}       & \multicolumn{1}{l|}{\begin{tabular}[c]{@{}l@{}}not\\ applicable\end{tabular}}                                                          & \multicolumn{1}{l|}{\begin{tabular}[c]{@{}l@{}}not\\ applicable\end{tabular}}        & \multicolumn{1}{l|}{\begin{tabular}[c]{@{}l@{}}not\\ applicable\end{tabular}}    & \multicolumn{1}{l|}{\begin{tabular}[c]{@{}l@{}}not\\ applicable\end{tabular}}     & {\color[HTML]{333333} \cmark}                \\ \hline
\multicolumn{1}{l|}{\cellcolor[HTML]{f4f4f4}\begin{tabular}[c]{@{}l@{}}is \textbf{AR headset}\\ sufficient?\end{tabular}}                     & \multicolumn{1}{l|}{\begin{tabular}[c]{@{}l@{}}+ markers,\\ foot pedals, mouse\end{tabular}} & \multicolumn{1}{l|}{\begin{tabular}[c]{@{}l@{}}Kinect\\ camera setup\end{tabular}}  & \multicolumn{1}{l|}{\begin{tabular}[c]{@{}l@{}}+ robotic\\ arm\end{tabular}}                                                           & \multicolumn{1}{l|}{\begin{tabular}[c]{@{}l@{}}+ pen,\\ tablet\end{tabular}}         & \multicolumn{1}{l|}{\cellcolor[HTML]{CCE3C3}{\color[HTML]{333333} \cmark}}          & \multicolumn{1}{l|}{tablet}                                                       & {\color[HTML]{333333} \cmark}                \\ \hline
\multicolumn{1}{l|}{\cellcolor[HTML]{f4f4f4}\begin{tabular}[c]{@{}l@{}}\textbf{open-source}\\ codebase\end{tabular}}                          & \multicolumn{1}{l|}{not provided}                                                            & \multicolumn{1}{l|}{not provided}                                             & \multicolumn{1}{l|}{not provided}                                                                                                      & \multicolumn{1}{l|}{not provided}                                                    & \multicolumn{1}{l|}{not provided}                                                & \multicolumn{1}{l|}{not provided}                                                 & {\color[HTML]{333333} \cmark}                \\ \hline
\end{tabular}%
}
\end{table*}

\subsection{XR in Digital Design and Fabrication}
XR technologies, which include virtual reality (VR), augmented reality (AR), and mixed reality (MR), have increasingly been applied to enhance various aspects of digital design and fabrication~\cite{dogan_fabricate_2022}.

\paragraph{Virtual Reality:} VR tools have been mostly developed for digital design compared to fabrication. Tools like \textit{Gravity Sketch} allow users to create 3D models in an immersive environment, providing a spatial understanding that traditional screen-based interfaces cannot offer. Systems such as \textit{VRSketchIn}~\cite{drey_vrsketchin_2020}, \textit{AdaptiBrush}~\cite{rosales_adaptibrush_2021}, \textit{HandPainter}~\cite{jiang_handpainter_2021}, and \textit{HPIPainting}~\cite{cai_hpipainting_2025} serve as interim sketching tools that can facilitate the design workflow. Researchers also studied the effectiveness of CAD tools in VR~\cite{bourdot_vrcad_2010, feeman_exploration_2018, cordeiro_survey_2019}. While these tools provide immersive spatial understanding for 3D modeling, they are typically limited to virtual-only environments, separating the design process from the physical context of deployment.

\paragraph{Augmented Reality:}
In contrast to VR, AR operates directly in physical environment, allowing users to see their designs in their intended real-world context~\cite{iyer_xr-penter_2025, dogangun_rampa_2024, dogan_standarone_2023}.
While head-mounted displays (HMDs) can be used with hand gestures (as seen in \systemName) or controllers in AR~\cite{guo_hololens_2022, stemasov_param_2024, peng_roma_2018, jasche_printarface_2020}, most of the previous AR systems rely on tablets with pen or finger interactions. For instance, \textit{massless}\footnote{\url{https://massless.dev/}, Accessed: 19.01.2025}, focuses on reducing material waste during prototyping by overlaying design information directly onto the artifact's deployment area, enhancing precision and minimizing errors due to physical constraints. \textit{Sketched Reality}~\cite{kaimoto_sketched_2022} introduces an AR sketching tool that interacts with physically actuated robots, exploring the physics of designs and mechanisms. \textit{RealitySketch}~\cite{suzuki_realitysketch_2020} introduces a dynamic and responsive interface for sketching interactive graphics and visualizations. 

\systemName contributes to the vision of works such as \cite{wang_pointshopar_2023, alghofaili_warpy_2023, leiva_rapido_2021, leiva_pronto_2020}, which streamline digital design and fabrication workflows. Some systems focus on specialized fabrication processes, such as \textit{Palette-PrintAR}~\cite{lipkowitz_palette-printar_2024}, which uses AR to design and simulate for multicolor resin 3D printing. Unlike these systems, \systemName uniquely integrates a 3D printing workflow directly within its AR interface, providing an all-in-one solution that eliminates the need for external slicing tools or connector software while supporting precise and fully-functional designs.

\paragraph{Mixed Reality:} James and Eckert~\cite{james_feasibility_2023} developed MR systems that allow operators to visualize potential outcomes projected onto the material via real-time physics simulation, enabling adjustments before a metal-cutting process begins. Other works explored conveying information spatially to users~\cite{chen_papertoplace_2023, cheng_interactionadapt_2023,  li_predicting_2024, han_blendmr_2023}.
These can be integrated with digital design and fabrication systems to provide instructions or context-aware interactable menus~\cite{dogan_augmented_2024, cheng_semanticadapt_2021, lindlbauer_context-aware_2019}
\textit{Window-Shaping}~\cite{huo_window-shaping_2017} enables rapid 3D model creation on and around physical objects using a sketch-and-inflate scheme using a tablet.

\vspace{0.1cm}

Building on this foundation, \systemName uniquely and efficiently integrates in-situ constructive solid geometry (CSG) modeling in a seamless 3D printing workflow within an immersive AR interface. Unlike existing systems, \systemName allows direct 3D design and printing in any workspace without requiring additional tools or post-processing software.

\subsection{In-Situ Tools and Interfaces}

In-situ tools refer to systems used directly within the environment where a task takes place, rather than in separate or abstract settings. Recent advancements in AR have enabled designers to visualize changes within the actual environment, bridging the gap between artifact creation and its intended deployment~\cite{ashbrook_towards_2016, mahapatra_barriers_2019}. For example, \textit{DesignAR}~\cite{reipschlager_designar_2019} combines 2D tablet screens with stereoscopic AR and a pen for precise object tracing. \textit{RoMA} synchronizes design and fabrication using a 3D printing robotic arm that incorporates existing objects into the workspace. \textit{HoloDesk}~\cite{hilliges_holodesk_2012} and \textit{MixFab}~\cite{weichel_mixfab_2014} introduce immersive tools with gestural controls tracked by \textit{Kinect}, enabling context-aware designs. Unlike systems requiring fixed setups or specialized hardware, \systemName allows design in and for \textit{any} environment using only an off-the-shelf HMD. This portability makes \systemName a versatile tool for diverse settings, whether at home or outside. %, broadening the scope of in-situ AR-based design.

The concept of in-situ is not only about overlaying information but also about interacting with the environment in a meaningful way. \textit{Teachable Reality}~\cite{monteiro_teachable_2023} lets users define their own in-situ tangible and gestural real-time interactions, allowing them to prototype functional AR applications without programming. \textit{SnapToReality}~\cite{nuernberger_snaptoreality_2016} enables real-time alignment of virtual objects to real-world constraints, such as edges and planar surfaces. \textit{FusePrint}~\cite{zhu_fuseprint_2016} explored design for fitting mounts, while \textit{Robiot}~\cite{li_robiot_2019} introduces designing of mechanisms that actuate other objects, and \textit{Protopiper}~\cite{agrawal_protopiper_2015} discusses room-scale design with low fidelity. %In relation, 
Mitterberger et al.~\cite{mitterberger_augmented_2020, mitterberger_augmented_2022, mitterberger_extended_2023} focus on larger-scale architectural designs. 

Some in-situ modeling systems follow a \textit{design in and for anywhere} ideology. Lau et al. introduced \textit{Modeling in Context}~\cite{lau_modeling--context_2010}, which uses everyday objects as a base point for designing around them. \textit{pARam}~\cite{stemasov_param_2024} provides in-situ parametric design, while \textit{Situated Modeling}~\cite{lau_situated_2012} utilizes fiducial markers for robust object tracking. 
While these markers can impact scene aesthetics and unobtrusiveness, researchers showed it is also possible to hide them using specialized imaging techniques~\cite{dogan_infraredtags_2022, dogan_brightmarker_2023}. 
\systemName builds on this \textit{design in and for anywhere} ideology but takes a novel direction by CSG modeling. Unlike sketching or parametric approaches, CSG modeling requires the integration of precise features such as snapping to a grid, object alignment, and robust control logic systems to enable functional and flexible design workflows. The adaptation of these traditionally desktop-bound features into an in-situ AR context is what allows \systemName to advance this ideology.

Tools incorporating tangible artifacts with in-situ design are valuable for context awareness and experimenting with physics, reducing the need for simulation. 
\textit{BrickStARt} used interlocking building blocks for this purpose~\cite{stemasov_brickstart_2023}.
\textit{Tangible Version Control}~\cite{letter_tangible_2022} explores versioning of physical artifacts with the help of an HMD. \textit{Printy3D}~\cite{yung_printy3d_2018} teaches 3D modeling with tangible artifacts placed in a virtual container.
Other tools include linkage~\cite{jeong_mechanism_2018} or carving~\cite{hattab_interactive_2019, hattab_rough_2019}, which make the design process interactive by touch.

Our system integrates tangibility into the design experience in several ways. It incorporates \textit{reference object types}, i.e., representations of object that serve as design references, equipped with functionalities to design around them. It also features \textit{depth occlusion} for \textit{physical} objects, providing a realistic interaction between the real and digital worlds.

\subsection{Novice Support Tools and Systems}

Designing systems for novices requires consideration of simplicity and ease of learning. In the context of digital design and fabrication, this often translates into reducing the complexity of traditional CAD tools and making them more accessible. \textit{Tinkercad}, for instance, has been successful in this regard by providing an entry-level CAD tool that embraces CSG modeling with simple geometric shapes and a drag-and-drop interface that is easy for beginners to grasp ~\cite{stemasov_road_2021}.
Other systems for design and fabrication using simplified core building blocks include \textit{FlatFitFab}~\cite{mccrae_flatfitfab_2014}, which connects planar parts for a 3D model, and similarly \textit{kyub}~\cite{baudisch_kyub_2019}, which uses simpler block-based design.
\textit{PARTs}~\cite{hofmann_greater_2018} addresses the challenges faced by non-experts in 3D modeling by introducing a framework for expressing and reusing design intent. \textit{CraftML}~\cite{yeh_craftml_2018} uses declarative syntax where the code dictates the output structure. Lee et al.~\cite{lee_interactive_2018, lee_posing_2016} use pose and movement detection for personalized furniture design, and Tian et al. simplify analog fabrication with woodcutter ~\cite{tian_matchsticks_2018} and lathe \cite{tian_turn-by-wire_2019} tools. In a similar vein, \textit{Draw2Cut}~\cite{gui_draw2cut_2025} lowers the barrier to CNC milling by allowing novices to design by sketching directly on the physical material.
Suzuki et al.~\cite{suzuki_augmented_2022} use virtual twins to visualize simulated behaviors when integrated with spatial references.
Pfeuffer et al.~\cite{pfeuffer_gaze_2017} shows the use of virtual buttons on the wrists to access interactable menus.

\begin{figure*}[ht]
  \centering
  \includegraphics[width=1\textwidth]{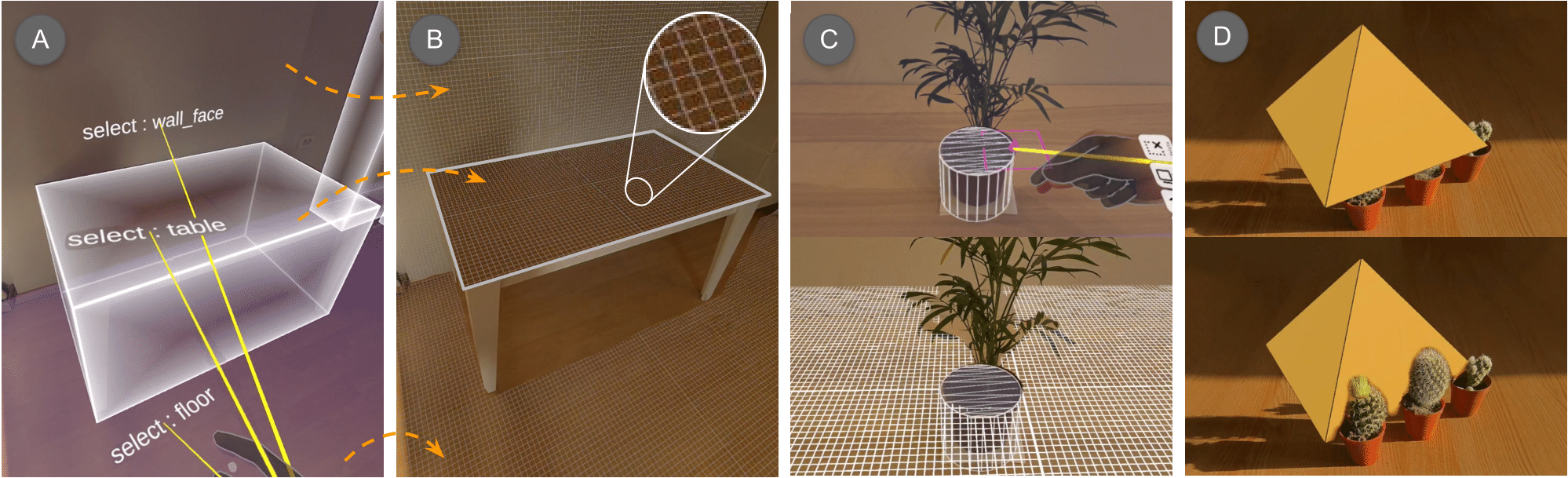}
  \caption{\textbf{Real-life workspace awareness}. (a) Three cases for selecting the workspace in a scanned room. (b) Grid appearing on the selected workspace plane for the three corresponding cases. (c) Selecting the workspace based on a reference of a physical object. (d) Physical and digital objects rendered together with depth occlusion off (upper) versus depth occlusion on (lower).}
  \Description{Four photos labeled (a), (b), and two in (c). The first two photos show the first-person view of the user looking at a room with a table inside. The other two photos show three small decorative cacti on a table with a digital pyramid model behind them. In the image (a), three yellow rays come out from the user's right hand, first hitting the floor where a text appearing above says "select: floor", second hitting the top of the table where a text appearing above says "select: table", and the third hitting the wall where a text appearing above says "select: wall". There is a white outline around the table. In image (b) the room is seen from the same angle as an image (a), but this time there are grids on top of the floor, table, and wall. In the first image of (c), the digital pyramid model partially blocks the view of the cacti even though it is placed behind them while in the second image of (c) the cacti are fully visible.}
  \label{fig:selecting_surfaces}
\end{figure*}

Generative design has also been instrumental in achieving desired 3D outcomes with minimal effort. Some tools allow users to design and fabricate 3D models based on 2D sketches \cite{saul_sketchchair_2010, li_sketch2cad_2020, kazi_dreamsketch_2017, johnson_sketch_2012, igarashi_teddy_1999}, thus reducing the complexity associated with directly creating models in 3D space. 
Other ways of generative design utilize speech inputs~\cite{ballagas_exploring_2019} and constrained parametric design~\cite{shugrina_fab_2015}.
Further, retrieving existing designs~\cite{stemasov_mixmatch_2020, liang_customizar_2022} and remixing~\cite{follmer_kidcad_2012, follmer_copycad_2010, stemasov_immersive_2023, roumen_grafter_2018} enhance accessibility while balancing expressivity \cite{stemasov_road_2021}. \textit{ShapeFindAR}~\cite{stemasov_shapefindar_2022} enables searching for physical artifacts through spatial queries and text.

Among novice-oriented tools, \systemName takes inspiration from \textit{Tinkercad}’s simplicity and accessibility, especially its use of CSG modeling. By adopting an approachable modeling paradigm like CSG, combining it with spatial interaction features, and integrating it into an end-to-end design and fabrication workflow, \systemName aims to lower the entry barrier for beginners and introduce new opportunities for creating designs directly in their context.

\section{DESIGN RATIONALE}
When developing \systemName, several key design considerations were prioritized to ensure the system was effective for novice users. The main considerations included are novice-friendly design and interaction, workspace awareness, and seamless integration:

\subsection{Novice-Friendly Design and Interaction}
To enhance the user experience in \systemName, we prioritize both simplicity in design and interaction. Our approach combines straightforward design logic with immersive user interactions to create an accessible and engaging environment for all users. The system’s design and interaction features are crafted to ensure that even novice users can seamlessly create and manipulate models, while advanced users benefit from an efficient design process.

\paragraph{Simplified Design for Novices:} \systemName is designed with a focus on simplicity to ensure a novice-friendly user experience. By utilizing \textbf{CSG} operations, users can easily create complex models by combining basic shapes (e.g., cuboids and spheres) through simple boolean operations such as union and subtraction. This approach reduces complexity, allowing users to design without managing technical parameters. The system categorizes shapes into \textbf{\textit{solid}} and \textbf{\textit{hole}} types, enabling the addition or removal of material to create more advanced designs effortlessly.

\paragraph{Natural Interaction:} Interactivity in \systemName is driven by \textbf{hand gestures} and \textbf{movement}, recognized by \textit{Quest}’s tracking system. These gestures allow users to manipulate objects directly. Building upon the methods outlined in prior works \cite{kim_tangible_2005, weichel_mixfab_2014, stemasov_param_2024}, \systemName lets users interact with the elements directly with their hands and without the need for external tools other than the headset. For instance, a pinching motion allows users to grab and move objects, while wrist rotation of the weak hand toggles features such as \textbf{snap-to-grid}, \textbf{multi-selection}, and \textbf{uniform scaling}. Moving the weak hand along global axes (x, y, z) toggles \textbf{axis and plane locks} for much needed precision while manipulating objects. The wrist-mounted menus provide an interface with buttons for quick navigation of available operations and settings. To increase design productivity, there are also shortcuts created that utilize pinch gestures with \textbf{middle and ring finger} for selection operations that do not require pointing at an object, which are the \textbf{selection and deselecting of all objects} created. Visual and auditory feedback confirms actions to further enhance the system's usability and create an immersive experience.

\subsection{Workspace Awareness} 
\systemName's real-world integration is achieved through sophisticated scene recognition and depth occlusion technologies. These allow the system to not only understand the environment but also intelligently interact with it.

\paragraph{Scene Recognition:} It is essential for the system to recognize and adapt to its physical surroundings by using real-time environment mapping to adjust designs according to available space and nearby objects. The system can identify furniture and other objects, categorizing them as potential design spaces or obstacles, such as a desk, wall, or floor. This recognition ensures accurate placement and scaling of models within the environment. For instance, the system's awareness of the selected workspace allows it to overlay a grid layout, enabling users to utilize snap-to-grid features and visualize object projections on the workspace as a reference during the design process. Additionally, the system supports the creation of \textit{\textbf{reference objects}}, which are simple wireframe geometries representing physical items (e.g., a plant pot when designing a plant pot holder). These reference objects enable users to place the workspace grid plane on them, facilitating designs that seamlessly integrate with real-world items.

\paragraph{Depth Occlusion:} It is also essential for users to understand their surroundings for context awareness while designing. While AR functionalities significantly aid in this area, models rendered in front of every world object can detract from the realistic nature we aim to achieve in the design task. By understanding the depth of objects in the environment, \systemName can occlude parts of a model that are behind physical objects. This enhances the realism and immersion of the design experience, while elevating user perception of the design as part of their real environment.

\subsection{Unified Design-to-Fabrication Workflow}
The transition between designing in AR and preparing a model for fabrication (in our case, 3D printing) should be fluid and straightforward, encouraging users to iterate on designs quickly and efficiently. 

\paragraph{Integrated Design and Fabrication:} In \systemName, we have consolidated design and fabrication functionalities into a single system, streamlining the entire workflow. Users can design their models and select print settings (e.g., layer height, infill pattern, support structure) directly within the AR environment, eliminating the need to transfer designs between separate tools. This simplifies the process by allowing users to handle all aspects of design and preparation within a unified interface, thus reducing the complexity traditionally associated with the use of multiple software packages.

\paragraph{Direct Communication:} Our system further enhances the user experience by enabling direct communication with WiFi-enabled 3D printers from within the system. This capability eliminates the need for a separate computing device and removes intermediate steps, such as exporting to slicing software, by sending designs directly to the printer.

\begin{figure*}[h]
  \centering
  \includegraphics[width=0.9\textwidth]{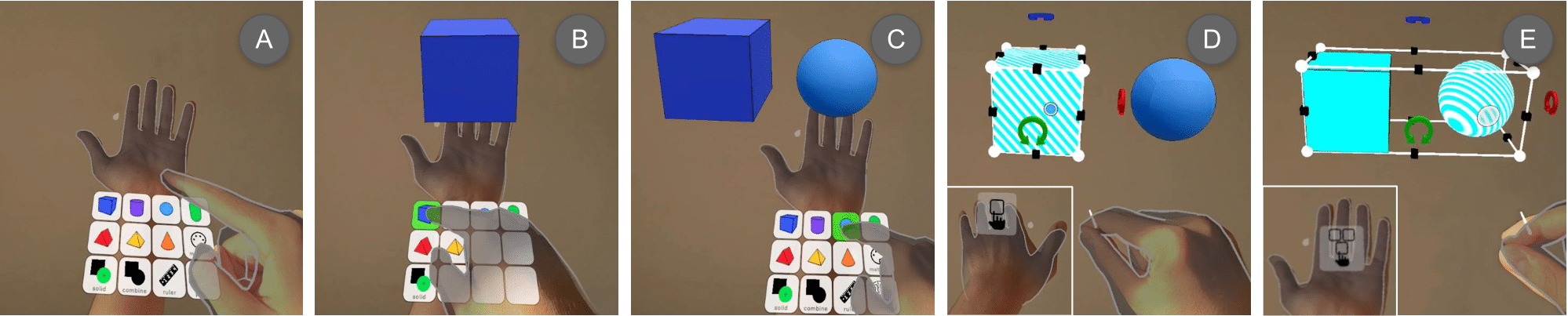}
  \caption{\textbf{Object creation and selection}. (a, b, c) Poking gesture on buttons for creating primitive objects. (d) Highlighting by pointing hand towards an object and selecting it by pinch move in single selection mode. (e) Selecting objects in multiple selection mode.}
  \Description{Five photos labeled (a), (b), (c), (d), (e) show the first-person view of a user in front of a wall interacting with their hand and the digital objects. In images (a), (b), and (c) user have extended their weak hand to their eye level and looking at the inside of their wrist. There are 12 white square buttons on the wrist in a three-by-four grid. Seven of the buttons show primitive geometric objects, namely cube, cylinder, sphere, capsule, triangular prism, pyramid, and cone. In the image (a), the user is hovering his dominant hand index finger on their wrist. In the image (b), the user pokes the cube button with their dominant hand index finger and the button turns green and a cube appears above the weak hand. In the image (c), the user does the same with a sphere next to the cube. In image (d), user does a pinch motion while their dominant hand hand is directed toward the created cube and their weak hand is oriented palm-down. The cube's color is changed from blue to cyan and a white grid is appearing on the cube. Similarly, in the image (e), the user does the same pinch motion for the sphere, this time with the weak hand oriented palm-up. Both of the objects turn to cyan, with white grid appearing on the sphere instead of the cube.}
  \label{fig:object_creation_and_selection}
\end{figure*}

\section{Implementation}

\systemName was implemented for \textit{Meta Quest 3}. The system code is written in C\# with \textit{Unity} Version \textit{2022.3.22f1 LTS}\footnote{\url{https://unity.com/releases/2022-lts}, Accessed: 19.07.2024}. The system utilizes open-source resources provided by \textit{Meta} to support various features. Specifically, we used \textit{Meta XR All-in-One SDK}\footnote{\url{https://developer.oculus.com/downloads/package/meta-xr-sdk-all-in-one-upm/}, Accessed: 19.07.2024}, which includes libraries for hand tracking, room and furniture detection (\textit{Scene API}\footnote{\url{https://developer.oculus.com/documentation/native/android/mobile-scene-api-ref/}, Accessed: 2.08.2024}), passthrough features, and other essential XR development tools for \textit{Unity}. Additionally, we incorporated \textit{Meta}'s \textit{Depth API}\footnote{\url{https://github.com/oculus-samples/Unity-DepthAPI}, Accessed: 19.07.2024} library for the depth occlusion feature and \textit{Unity}'s built-in audio engine for the spatial audio utilities.
The subsequent sections will outline the envisioned procedural steps of using \systemName with its features, from (1) selecting the workspace to (2) designing artifacts, and finally to (3) fabrication of those artifacts.

\subsection{Workspace Selection and Modification}
\systemName system enables users to select their desired workspace, which can vary from the top of a desk to the floor, walls, door frames, and other flat surfaces.

\subsubsection{Scene Setup} First, users scan their rooms using the headset's built-in space setup feature. This feature identifies the room's boundaries and furniture while providing users the option to manually configure their furniture (e.g., for selecting which part of the desk they would like to work on). The scanned furniture is saved as rectangular prism models, representing all selectable workspaces as simple, flat surfaces (\autoref{fig:selecting_surfaces}a). 

\subsubsection{Workspace Selection} After scanning, users can interact in \systemName and view the highlighted selectable workspaces. This is possible with the headset's built-in room recognition feature and \textit{MRUK} framework\footnote{\url{https://developer.oculus.com/documentation/unity/unity-mr-utility-kit-overview/}, Accessed: 20.07.2024}. Then, users use their dominant hand to direct a ray toward the surfaces of furniture or room boundaries, which displays the name of the planned workspace it intersects (\autoref{fig:selecting_surfaces}a). An index finger pinch gesture is then employed to confirm the selection.

\subsubsection{Workspace Modification} After the workspace is selected, a grid is overlaid on the chosen surface, covering the entire area of the furniture's selected side or the boundary (\autoref{fig:selecting_surfaces}b). This grid facilitates snap-to-grid operations, enhancing precision in the design process. While this grid is visually planar, its snap points extend alongside the grid plane's normal direction, spreading snap points into the scene space in all dimensions. The density of the lines on the grid depends on the grid spacing and by default, the grid spacing is set to 2 cm, with the option to adjust this value via the hand menus. Additional grid options include selecting its color, which enhances visibility (e.g., a white grid on a white surface may be less visible), determining whether physical objects should occlude the grid, and adjusting its distance to the surface alongside its normal, which can be useful against z-fighting issues when the grid is in occlusion mode. Additionally, depth occlusion for digital objects can be toggled on and off (\autoref{fig:selecting_surfaces}c), aiding in designing with physical world constraints and designing artifacts that interact with real-world objects.

\subsection{Design Features}
Once users select their workspace, they can begin designing models directly with their hands, leveraging an array of tools typically found in traditional CAD systems.

\subsubsection{Object Creation}
Users can choose from seven types of primitive 3D geometries (cube, sphere, cylinder, capsule, triangular prism, pyramid, and cone) using buttons on their weak hand wrist (\autoref{fig:object_creation_and_selection}a). The color of the button turns blue when the finger hovers close to the button. This gives the user feedback on which button is about to be pressed. Upon pressing a button, the selected object appears on the user's weak hand (\autoref{fig:object_creation_and_selection}b and \ref{fig:object_creation_and_selection}c).

\subsubsection{Object Highlighting and Selection}
During interactions, users may encounter multiple instances of objects, necessitating methods for selecting and manipulating specific items.

\vspace{-0.2cm}

\paragraph{Highlighting} To select an object, users point their hand toward it, with an invisible ray cast from the tracked hand model in the direction of the pointing finger. When this ray intersects with an object, visual feedback is provided (\autoref{fig:object_creation_and_selection}d). Highlighted objects are rendered with a white grid overlay featuring transparent lines, ensuring the object’s original color remains visible.

\vspace{-0.2cm}

\paragraph{Selection} 
A pinch gesture, i.e., thumb and index finger touching, selects a highlighted object (\autoref{fig:object_creation_and_selection}d), turning it bright cyan and emitting a directional sound. A surrounding \textit{Object Manipulation Box} appears, enabling movement, resizing, and rotation.

\subsubsection{Single and Multiple Selection}
Selection modes toggle based on weak-hand orientation: palm down for single selection (default, \autoref{fig:object_creation_and_selection}d) and palm up for multiple selection (\autoref{fig:object_creation_and_selection}e).

\vspace{-0.2cm}

\paragraph{Single Selection} 
In the single selection mode, only one object is active; selecting another deselects the current object.
As this is considered the default mode, the palm-down orientation is chosen as it is the resting position of the hand.

\vspace{-0.2cm}

\paragraph{Multiple Selection} 
This mode enables group manipulation, i.e., the equivalent of pressing the \textit{control} key on a keyboard in desktop interfaces.
Selecting an object that is not already selected adds it to the list of selected objects. Conversely, selecting an already selected object removes it from the list.
Multiple selected objects can be manipulated together at once.

\subsubsection{Object Manipulation} 
The \textit{Object Manipulation Box} facilitates object movement, resizing, and rotation.

\vspace{-0.2cm}
\begin{figure*}[h]
  \centering
  \includegraphics[width=1\textwidth]{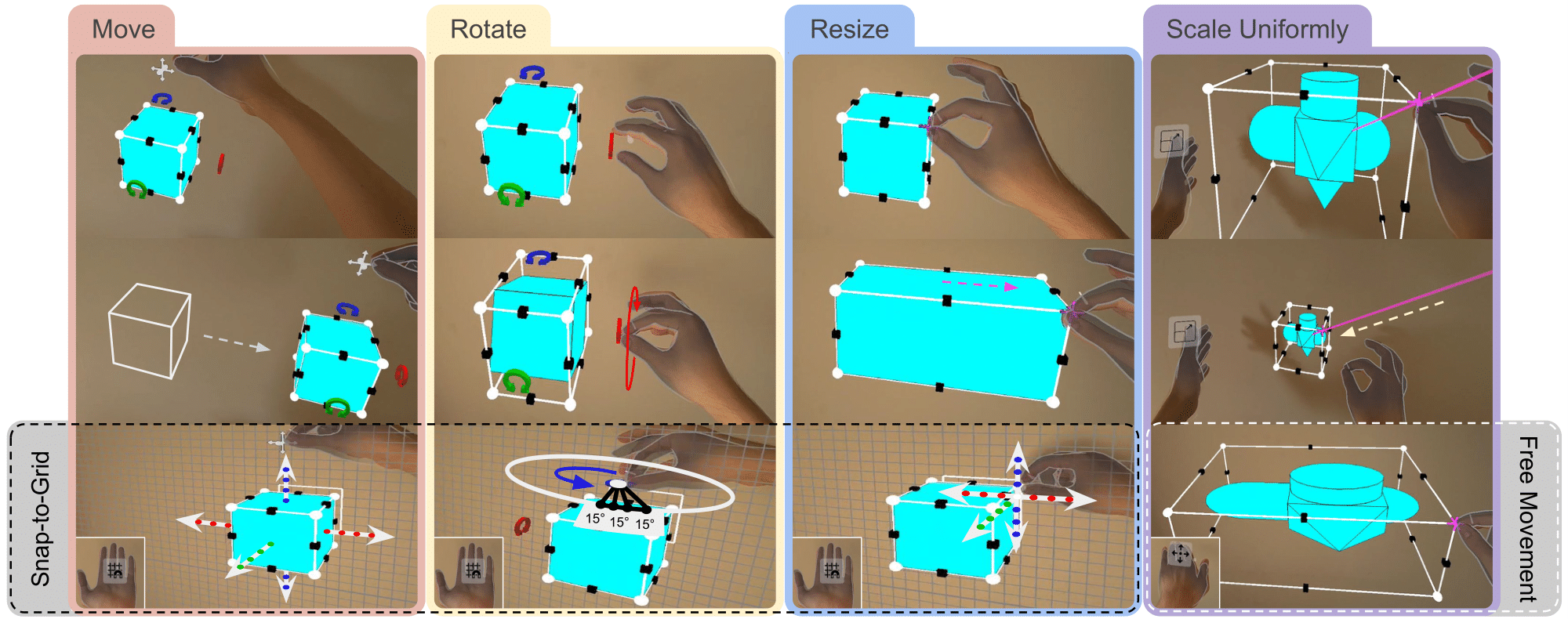}
  \caption{\textbf{Different ways of manipulating objects}. Users can perform the following main actions: \textbf{Moving} objects around the world space with the movement handle; \textbf{rotating} objects by twisting the rotation handle around its corresponding axis; and \textbf{resizing} objects by dragging the corner handles. The orientation of the weak-hand toggles manipulation modes between \textbf{free movement} (palm-down), \textbf{uniform scaling} (palm-sideways), and \textbf{snapping-to-grid} (palm-up).}
  \Description{Six photos labeled (a), (b), (c) each label corresponding to two photos show a cube being manipulated with their dominant hand from the first-person view of the user in front of a wall. The first pair of pictures labeled (a) shows the user pinching the movement handle on top of the cube in the first picture and dragging their dominant hand and the object to the right in the second picture. The second pair of pictures labeled (b) shows the user pinching the rotation handle next to the cube in the first picture and rotating their dominant hand wrist and the cube in the second picture. The third pair of pictures labeled (c) shows the user pinching the corner handle at the top right corner of the cube in the first picture and moving their hand to the right and the cube stretching to the right in the second picture.}
  \label{fig:objectmovement_rotation_resizing}
\end{figure*}

\paragraph{Moving Objects} 
A movement handle, i.e., gizmo, above the objects follows the user’s pinch gesture, which allows for freehand movement (\autoref{fig:objectmovement_rotation_resizing}).

\vspace{-0.2cm}

\paragraph{Rotating Objects} 
Three axis-aligned rotation handles allow rotation around the x, y, and z axes. Grasping a handle with a pinch gesture rotates objects like a knob (\autoref{fig:objectmovement_rotation_resizing}).

\vspace{-0.2cm}

\paragraph{Resizing Objects} 
Corner handles at each vertex of the \textit{Object Manipulation Box} enable resizing. Adjusting a handle moves it freely while the opposite handle remains fixed, which ensures precise control (\autoref{fig:objectmovement_rotation_resizing}). 

\vspace{-0.2cm}

\paragraph{Uniform scaling} Gets activated when the weak hand is sideways (\autoref{fig:objectmovement_rotation_resizing}). A guiding line extends from the box center through the selected corner, ensuring accurate scaling along this line. Multiple selections allow synchronized scaling, while free scaling does not preserve aspect ratios.

%\begin{figure*}[h]
%  \centering
%  \includegraphics[width=0.95\textwidth]{figures/NonUniformvsUniformScaling.pdf}
%  \caption{\textbf{Uniform vs. non-uniform scaling}. (a, b) Resizing objects with uniform scaling mode. (c) In free scaling mode.}
%  \Description{Three photos labeled (a), (b), and (c) show a first-person view of a user looking at a wall and interacting with a complex 3D abstract design. In images (a) and (b), the user has their weak hand's palm looking right and they apply a pinching motion to the corner handle on the upper right side of the manipulation box. A pink ray is seen coming from the middle and going through the selected corner handle. The user brings their right hand in the second picture closer to the center of the manipulation box, resulting in a smaller object with the same aspect ratio. Image (c) shows the user dragging the same corner handle with their weak hand palm looking down. The resulting object is compressed vertically.}
%  \label{fig:uniform_scaling}
%\end{figure*}

\vspace{-0.2cm}

\paragraph{Snap-to-Grid} When any of the elements for manipulating objects are held with the dominant hand with a pinch gesture, the orientation of the weak hand dictates the toggle between free movement and snap-to-grid modes instead of single and multiple selection modes. Free movement is enabled when the weak hand's palm faces upward, while snap-to-grid is activated when the palm faces downward. The 
functionality adapts to movement, rotation, and scaling operations (\autoref{fig:objectmovement_rotation_resizing}).

\textit{For movement}, when the snap-to-grid feature is activated, the \textit{Object Manipulation Box} will align with the grid lines on the selected workspace. In some cases, both parallel sides of an object may not be able to snap to the grid lines due to the distance between them. In such situations, the system determines which side should snap to a grid line based on the following logic:

\begin{enumerate}[leftmargin=0.8cm]
    \item If a side is parallel to the grid plane, the system will always favor the side closer to the grid. (e.g., placing the bottom of the object on the table grid)
    \item If a side is not parallel to the grid plane, the system will favor the side that aligns with the direction of movement calculated from the start location of the grabbing action.
\end{enumerate}

\begin{figure}[b]
  \centering
  \includegraphics[width=0.47\textwidth]{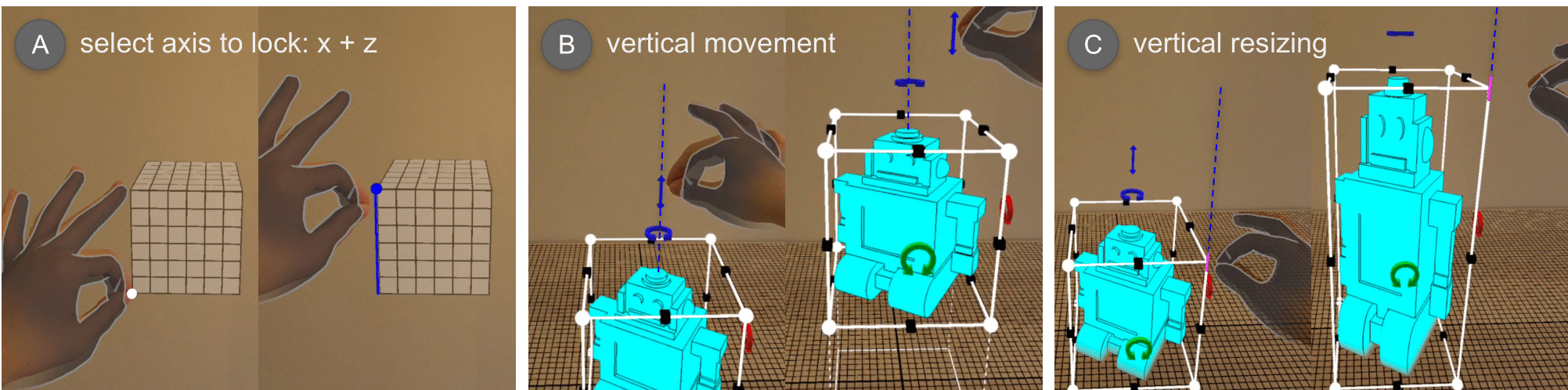}
  \caption{\textbf{Axis locking}. Users can (a) initiate the axis lock by pinching with their weak hand and moving it in the desired direction. This helps (b) restrict the  movement and (c) resize the objects.}
  \Description{}
  \label{fig:axis_lock}
\end{figure}

\textit{For rotation}, when the snapping option is enabled, the rotation handle aligns with 15-degree increments from its initial position by default. Smaller increment values allow for greater rotational flexibility, though they may reduce precision in achieving the desired orientation due to increased likelihood of overshooting or undershooting, especially at finer resolutions such as 1-degree steps.

\textit{For scaling}, when the snap-to-grid feature is activated, the corner handle aligns with the nearest grid point within the scene space to the pinch location. The objects are scaled in the same logic as in the free-motion mode, where the opposite corner acts as an anchor.
We note that the \textit{cell size} of the grid for movement and scaling, as well as the \textit{snapping degrees} for rotation are adjustable through settings.

\begin{figure*}[h]
  \centering
  \includegraphics[width=0.93\textwidth]{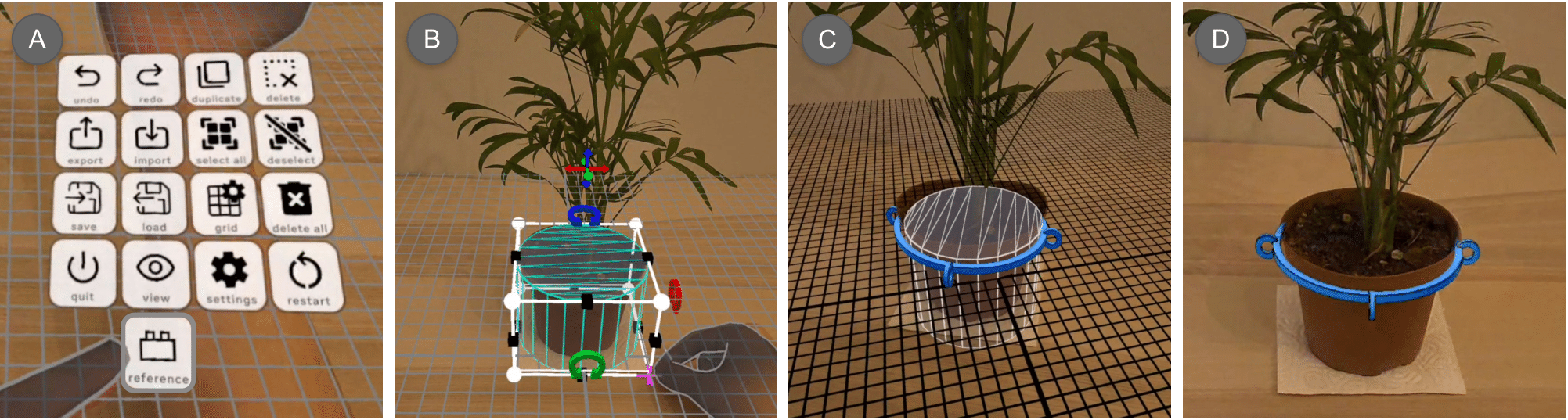} 
  \caption{\textbf{Reference objects.} Users can (a) create reference objects via a button and (b) use them to represent physical items. (c) On-object grids can be created for alignment on the reference object for an object-focused design. (d) The example shows a plant pot hanger designed around a reference pot.}
  \Description{}
  \label{fig:reference_object}
\end{figure*}

\subsubsection{Axis Locking}
The \textit{axis lock tool} in \systemName allows users to easily constrain object manipulation to specific axes or planes. While this is a fundamental operation, our approach to selection logic sets this mechanism apart from others. Users can activate the axis lock tool by performing an index finger pinch gesture with their weak hand. Moving their hand in the desired global direction triggers a visual cue in the form of a colored edge (\autoref{fig:axis_lock}a) or a side (if moved diagonally) of the \textit{axis lock tool}, representing the \textit{active axis or plane} for manipulation. This interaction ensures that users can limit movement or resizing action along a single axis or within a plane (\autoref{fig:axis_lock}b and \ref{fig:axis_lock}c) without needing to interact with additional buttons or menus.

A single tap using the pinch gesture with the weak hand resets the axis lock and can start moving in all directions again. We chose this to allow users to resume unconstrained manipulation quickly. In the development phase, we tested axis-specific handles around the \textit{Object Manipulation Box}, but it was apparent that these handles drastically crowded the area around the selected objects, which made the objects harder to see and interact with.

\subsubsection{Solid and Hole Objects}
Users can toggle whether an object is \textit{solid}, adding material, or \textit{hole}, removing material, by pressing a button on their weak hand wrist. The hole objects appear with a transparent mask, making them identifiable when compared to the solid objects (\autoref{fig:solid_and_hole}a). The hole objects can be manipulated the same way as the solid objects.

\subsubsection{Combining Objects}
\begin{figure}[b]
  \centering
  \includegraphics[width=0.47\textwidth]{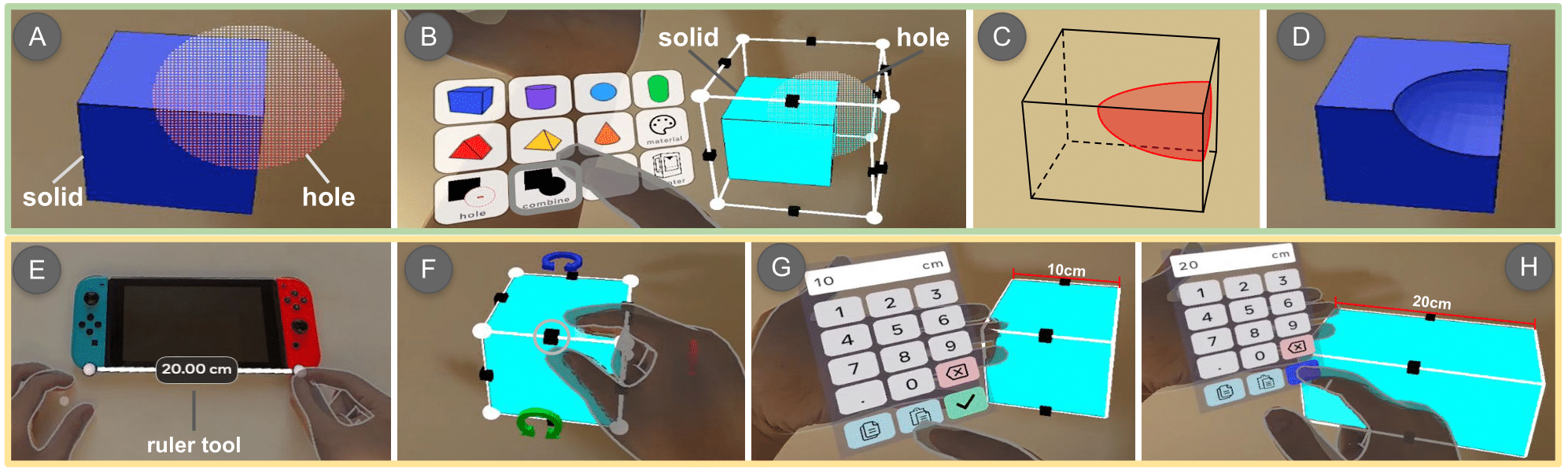}
  \caption{\textbf{Multi-object operations and in-context measurement}. Users can create (a) solid and hole objects. (b, c, d) The combining operation creates a cut in the intersection between the solid and hole object. Users can (e) measure the length of the objects around them with the ruler tool. For parametric resizing, users (f) select the handles on the edges of the object manipulation box, which (g) activates a numerical keyboard on the weak hand for (h) inputting the desired length alongside the edges.}
  \Description{Three photos labeled (a), (b), and (c) show a first-person view of a user looking at a wall. In the image (a) a solid cube and a hole sphere intersecting with the top corner of the cube is seen. The hole (or negative) sphere is seen with red color and with dotted transparent skin. In image (b) the weak hand wrist is visible while the cube and the sphere are selected and colored cyan. The user is about to poke a button named "combine" with their dominant-hand index finger. In the image (c), the combination of the objects is visible where there is a round cut on the upper corner of the cube where the sphere used to intersect the cube.}
  \label{fig:solid_and_hole}
\end{figure}

Users can combine and turn selected objects into one uniform object with a button on the weak hand wrist. The combined object will turn into the color of the last solid object chosen. The combining operation works with solid and hole objects, removing the hole objects after performing a cut on the solid objects through the intersection between the solid and hole objects (\autoref{fig:solid_and_hole}b-d).

\subsubsection{Ruler Tool}
Users can create a virtual ruler by pressing a button on their weak hand wrist. This simple ruler, equipped with two interactable edge handles (similar to scaling handles), provides real-world scale measurements. This ruler can be used for precise alignment and measuring distances between objects within the real and virtual space (\autoref{fig:solid_and_hole}e).

%\begin{figure*}[h]
%  \centering
%  \includegraphics[width=0.95\textwidth]{figures/UpdatedRulerTool.pdf} 
%  \caption{\textbf{In-context measurement}. Users can (a) measure the length of the objects around them with the ruler tool. For parametric resizing, users (b) select the handles on the edges of the object manipulation box, which (c) activates a numerical keyboard on the weak hand for (d) inputting the desired length alongside the edges.}
%  \Description{Four images labeled (a), (b), (c), and (d) show a user's first-person view using the system. In picture (a), user is using their weak and dominant hand to pinch both ends of a digital line where in the middle of the line it says "20.00 cm". The user is measuring out the length of a game console (Nintendo Switch). In pictures (b), (c), and (d) user is manipulating a digital cube model. In the picture (b) user is applying a pinch motion to a small black handle on the middle of the edge of the cube. In picture (c), a numerical keyboard interface appears on the weak hand in direction of the palm and it says "10 cm" in the input field of the keyboard interface. In picture (d), the user has inputted "20 cm" as the input and clicks the confirmation button. The cube's width is doubled.}
%  \label{fig:parametric_scaling}
%\end{figure*}

\subsubsection{Parametric Resizing}
In addition to resizing with free movement, snapping-to-grid, and uniform scaling operations, users can also adjust object sizes through precise numerical length inputs. To use this feature, users first select the edge of the \textit{Object Manipulation Box} they wish to resize (\autoref{fig:solid_and_hole}f) using a pinch gesture. Then, a numerical keyboard appears on the user's weak hand (\autoref{fig:solid_and_hole}g), allowing them to enter the desired real-world length for the selected edge. The object scales proportionally along the direction of the chosen edge based on this input (\autoref{fig:solid_and_hole}h). This method can be effectively combined with the ruler tool to ensure accurate design according to specific measurements.

\begin{figure*}[h]
  \centering
  \includegraphics[width=0.95\textwidth]{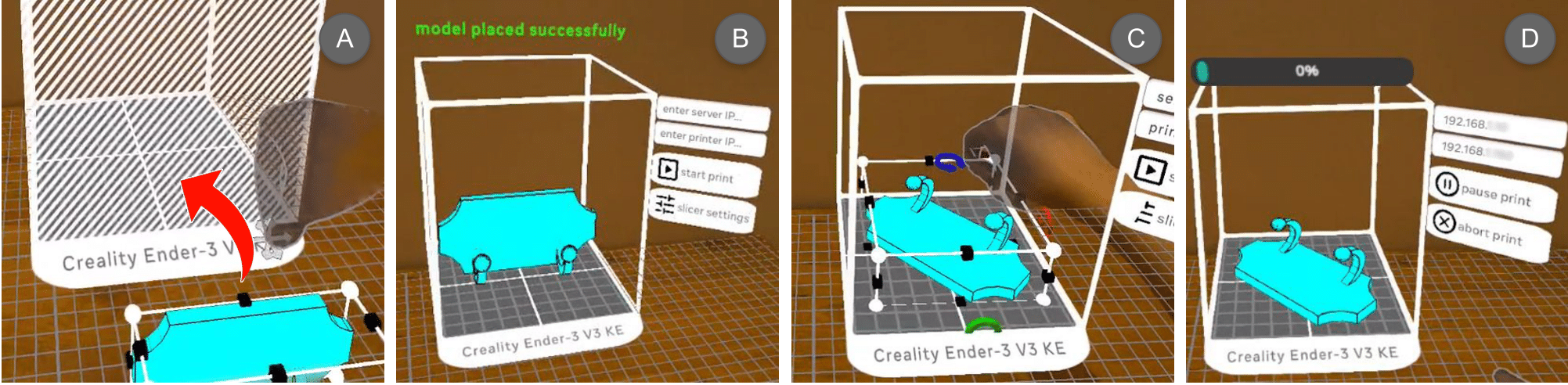}
  \caption{\textbf{Transition to fabrication.} Virtual 3D printer twin is used as a placeholder for exporting the model for a target printer or conducting a quick size check. Users can (a, b) drag and drop the models they create in the printer, (c) modify the print orientation, and (d) instantiate physical printing.}
  \Description{Four images labeled (a), (b), (c), and (d) show a user's first-person view looking on top of a table where there is a horizontal grid appearing on the table and a virtual 3D printer twin and a model of a wall hook is visible. The 3D printer twin is seen as a transparent rectangular prism with white edges and a grid on the bottom plate. The name of the printer (Creality Ender-3 V3 KE) is written on the bottom of the printer. In the picture (a), the user is moving the model with the movement handle grabbed with a pinch gesture with the dominant hand, dragging it to the printer twin. A white line grid appears on the printer twin. In picture (b), the model is placed in the middle of the printer twin and it says "model printed successfully" in green color on top of the printer twin. On the right side of the printer twin, two input fields and two buttons are visible. The first input field is labeled "enter sever IP..." and the second input field is labeled  "enter printer IP...". The first button is labeled "Start print" and the second button is labeled "Slicer settings". In the picture (c), user is rotating the model with a pinch gesture applied to the rotation handle on top of the model. In picture (d), a progress bar is seen on top of the printer twin and a text is written as "0\%" in the middle. The input fields contain the partially blurred IP addresses of the server and the printer and the buttons are labeled "pause print" and "abort print".}
  \label{fig:printer_twin}
\end{figure*}

\subsubsection{Reference Objects}
\systemName supports creating reference objects, offering users a powerful tool for designing around physical items. Users can activate the \textit{reference mode} via the hand menu on their dominant hand wrist (\autoref{fig:reference_object}a), which converts default block objects into reference objects. These reference objects are transparent and feature highlighted wireframe edges, distinguishing them from standard objects.

This wireframe structure allows users to precisely position grids on reference objects, aligning them with vertices and plane normals for accurate design. This translates to the grid not only functioning horizontally or vertically (as seen on the common workspaces) but also diagonally in any direction, fitting to the selected surface of the reference object.  The transparency aids in manipulating these objects to replicate physical counterparts (\autoref{fig:reference_object}b,c), enabling users to build around them easily. Additionally, reference objects can be set as \textit{hole} types, letting users capture intricate details of physical objects for better integration.

\subsubsection{Other Functions}
Other buttons placed on the dominant hand wrist enhance the efficiency of the design process. These features include selecting the color and material of the objects, duplicating selected objects, selecting or deselecting all objects, deleting selected objects, deleting all objects, undo, redo, saving, and loading operations.

To streamline the selection process, pinch gesture shortcuts were added. A pinch between the thumb and the middle finger triggers the function to deselect all objects, while a pinch between the thumb and the ring finger triggers the function to select all objects.

\subsection{Fabrication Features}
After the target model has been created, the user can export it to an STL file. This will allow users to access it from other CAD tools and enable them to 3D print their designs. This feature not only accelerates and simplifies the fabrication process but also enables users to quickly fabricate a draft of their design to assess whether it will be printed with the fidelity they anticipate.

The implementation of the server system was possible with the usage of \textit{Docker} containers to package the working environment for different computers. The server itself was written with \textit{JavaScript} in \textit{Node.js}, and the slicer used for the server was \textit{Ultimaker CuraEngine}\footnote{\url{https://ultimaker.com/software/ultimaker-cura/}}.

\begin{figure*}[h]
  \centering
  \includegraphics[width=0.9\textwidth]{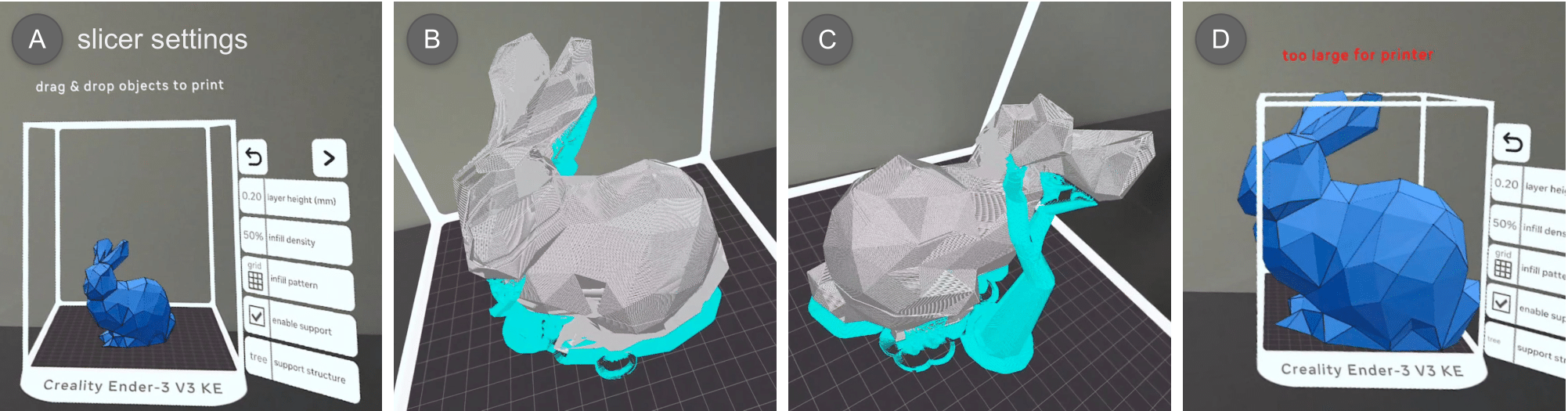}
  \caption{\textbf{Preparing for 3D printing and slicing.} (a) Users can select the slicing options for their print. They initiate slicing (b) and (c) view the sliced preview of the model, with outer walls in gray and support structures in cyan. (d) If a model exceeds the 3D printer’s build area, a "\textit{too large for printer}" indicator appears above the printer twin.}
  \Description{The low-poly Stanford bunny is placed inside the virtual printer twin with slicing options displayed next to it.}
  \label{fig:printer_slicing}
\end{figure*}

\subsubsection{Spatial Virtual Twin of 3D Printer}
A printer twin is a unique object that cannot be selected in the multiple selection mode or cannot be merged with other objects. This virtual printer object helps the user estimate how the object will be printed inside the real-world 3D printer. The size of the virtual printer twin corresponds to the real-world 3D printer's build area. This virtual printer's size can be selected from the predefined printers or set manually. To start the 3D printing process, the user has to drag and drop the current model into the printer twin. If neither the server IP nor the printer IP is given on the printer twin, it will save the current model locally into an STL file. If the slicer server is connected to the headset, the user can send their STL files to the server to create a G-code file for 3D printers. If the printer is also connected to the server, then the 3D printing process will begin automatically. 

\subsubsection{Drag and Drop and Adjustments on Build Plane}
Users can pick up the models they created, and drag them to the virtual 3D printer twin with the movement handle, where the printer will be highlighted (\autoref{fig:printer_twin}a). After a model is dropped over the printer twin, the duplicate of the objects comprised by the model will be placed inside the printer twin (\autoref{fig:printer_twin}b). The original objects will be moved to their initial position. After the object is placed inside the printer twin, users can adjust the size, orientation, and position of the object (\autoref{fig:printer_twin}c). Note that a print can not be initiated when the objects go outside the boundaries of the build area.

\subsubsection{Preparing for 3D Printing and Slicing}

The sliced model is visualized in the virtual 3D printer (\autoref{fig:printer_slicing}b and \ref{fig:printer_slicing}c), where the main walls are displayed in gray and support structures in cyan (tree support selected). The visualization allows users to review the slicing results and identify areas requiring adjustment before starting the print. If the selected model exceeds the 3D printer’s build area, the system notifies users by displaying a "\textit{too large for printer}" indicator above the virtual printer twin (\autoref{fig:printer_slicing}d). This ensures that users can modify their design or scale it appropriately before initiating the print.

After the printing request is sent through the buttons next to the virtual 3D printer twin object, if no server IP is given, the STL model of the object in the virtual printer will be saved locally. If the server IP is given while the printer IP is not, this STL model and sliced G-code file will be saved on the remote server. However, if both IPs are given, the printing process will start after slicing. The headset monitors those steps. The server IP is the same as the computer the server program is running on, and the printer IP can be obtained from the printer after it has been started and connected to the internet. Note that the printer should be accessible from the server computer if they do not reside on the same subnet.

The server manages multiple slicers to serve the XR headset clients. An example slicer configuration for testing has been created for \textit{Ultimaker S3}\footnote{\url{https://ultimaker.com/3d-printers/s-series/ultimaker-s3/}} printer with options:
{\small\texttt{Enable Support}}, 
{\small\texttt{Support Structure}}, 
{\small\texttt{Layer Height}}, 
{\small\texttt{Infill Density}}, and 
{\small\texttt{Infill Pattern}} (\autoref{fig:printer_slicing}a). 
More slicers can be implemented by adding them to the {\small\texttt{slicer\_options.json}} file. The slicer options are completely modifiable and could be increased with ease as the run command is also read from the JSON. How the slicer should work is completely in the hands of the programmer as the command could be easily manipulated by the flexible replacement tags that have been noted as options. The only required arguments that a slicer should take are output and input file arguments, which use {\small\texttt{gcode\_path}} and {\small\texttt{stl\_path}}, respectively.

A client can make multiple requests on the server. When a {\small\texttt{/slice}} GET request is sent to the server, it returns the available slicers and their options. On the other hand, when a {\small\texttt{/slice}} POST request is sent to the server, the STL file is loaded on the appropriate slicer, and the G-code file of the target model is obtained. The server parses the options selected by the user and passes them to the slicer. After the slicing operation has been completed, the resulting G-code file is visualized in the virtual 3D printer. This draft object is shown in different colors for the main and the support parts. After that, the 3D printing process can start. Note that one server can handle multiple printers, and one headset can connect to various servers. The only necessary action is to enter IP addresses.

\subsubsection{3D Printing and Monitoring}
After the {\small\texttt{/print}} POST request is sent to the virtual printer twin object, the server will connect and get access permission to the 3D printer. This permission is only necessary the first time as the server will save the token for accessing the printer, and access permission will not be necessary until the token has expired. If access to the 3D printer is granted, the server will send a printing request with the G-code file to the printer. The headset can obtain the print process information by sending the {\small\texttt{/print}} GET request. The user can pause or abort the printing process by sending a PUT request to the {\small\texttt{/print}} URL if necessary. This request supports three different commands: continue, pause, and stop, which can be sent via the buttons next to the 3D printer twin (\autoref{fig:printer_twin}d).
\section{Applications}
\label{sec:applications}

\begin{figure*}[h]
  \centering
  \includegraphics[width=0.97\textwidth]{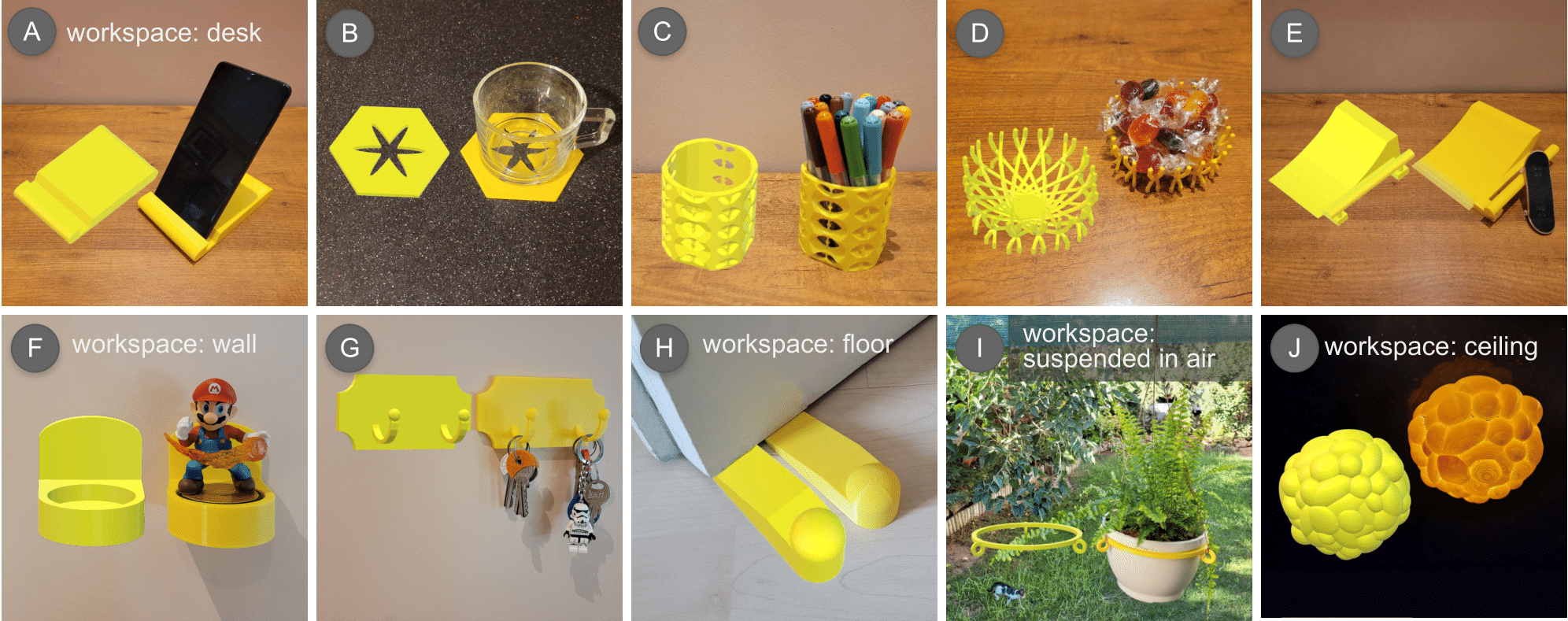}
  \caption{Designs created via \systemName on the left; their fabricated results on the right, placed in their intended space. These are (a) phone holder, (b) coaster, (c) pen holder, (d) candy bowl, (e) finger skateboard obstacle, (f) figurine stand, (g) key hanger, (h) door wedge, (i) plant pot hanger, (j) lamp shade.}
  \Description{Ten images labeled (a), (b), (c), (d), (e), (f), (g), (h), (i), and (j) show ten different 3D printed results next to their virtual rendered versions. The image (a) has a phone holder with a smartphone on top. In the image (b), there is a drink coaster with a cup on top of it. The image (c) has a pen holder with colorful markers inside it. In the image (d), there is a candy bowl with packaged candy inside. In the image (e), there is a ramp and a grinding rail with a finger skateboard next to it. In image (f), a figurine stand is mounted to a wall with a small figurine. In the image (g), there is a wall key hanger with two hooks and a bunch of keys hanging down those hooks. In the image (h), there is a door wedge holding a door from closing. In image (i), there is a plant pot carrying a ring connected to ropes that are hanging from the ceiling. In the image (j), there is a spherical light bulb shade covering a light bulb mounted to the ceiling.}
  \label{fig:applications}
\end{figure*}

In this section, we present ten distinct designs created with \systemName (\autoref{fig:applications}) to demonstrate its versatility and capabilities across various workspace environments. These examples highlight how \systemName enables users to easily produce both practical and creative objects through spatial interaction, tailored to their immediate surroundings.

\subsection{Desk Workspace} In the desk workspace, five designs were developed: a phone stand, drink coaster, pen holder, candy bowl, and fidget terrain (\autoref{fig:applications}a-e). These objects demonstrate the system's ability to facilitate personalized designs suited for everyday use. The phone stand and drink coaster highlight basic cutting operations and represent simple designs in everyday objects. The pen holder showcases intricate decorative patterns, made by combining and duplicating \textit{hole} objects in a chain, and the candy bowl explores experimental shapes by combining rectangular prisms into curvilinear forms. Fidget terrain stands out as a playful interactive object designed with ramps and rails, which illustrates the potential for creating designs that interact with moving physical objects. Each design benefits from real-time in-situ customization, ensuring the objects fit and function precisely within the user's workspace. Here, working on smaller scales with hand interactions can be challenging; thus, a useful approach is to create the design on a larger scale and then scale it down to the intended size. This helps manage precision and detail when working with smaller components.

\subsection{Wall Workspace} For the wall workspace, two objects were designed: a figurine stand and a key hanger (\autoref{fig:applications}f and \ref{fig:applications}g). 
The figurine stand allows users to display small decorative items without adding clutter, created using the ruler tool for a fitting hole. The key hanger offers practical functionality with aligned hooks using snap-to-grid. The models are snapped to the wall while designing for a better sense of depth, helping users create visually integrated solutions. 

\subsection{Other Workspaces} Beyond the more common desk and wall spaces, \systemName can also be applied to other environments, e.g., ground, ceiling, doors, windows, screens, couches, and more. On the ground, a door wedge was designed to hold doors securely in place (\autoref{fig:applications}h), benefiting from real-time scaling for accurate size. It was also crucial to have a feature where users can work around and build with existing objects in mind, so the coordinate system does not only depend on flat surfaces to stand on, but also considers the real-world objects they will interact with. The reference objects and the ability to modify the coordinate system based on any vertex of a reference object are directed towards this solution.

For the ceiling, we created two distinct designs: a plant pot hanger and a light shade (\autoref{fig:applications}i and \ref{fig:applications}j). The plant pot hanger demonstrates the system's ability to accommodate objects that require support and suspension, such as using "ropes" for hanging, a practical solution for overhead applications.
The light shade, on the other hand, highlights the system's capability to handle complex and abstract 3D modeling. Its artistic design illustrates the system's flexibility in creating intricate structures on a headset without technical limitations.
These examples show the adaptability to less conventional design spaces and orientations.

\section{Evaluation}
\label{user-study}
To evaluate \systemName, we conducted a user study involving three design tasks of varying contexts, comparing to \textit{Tinkercad} as a baseline. We used both quantitative measures and qualitative feedback to gauge user performance.

\subsection{Methods}

\paragraph{Participants}
We recruited ten participants (ages 23-29; 9 male, 1 female).
Four participants had little experience with CAD software, and six had no previous CAD experience.
Additionally, three participants had little experience with AR systems, and seven had no experience with AR systems.
Finally, four have used 3D printers in some parts of their life, while six have never used a 3D printer.

\begin{figure*}[h]
  \centering
  \includegraphics[width=1\textwidth]{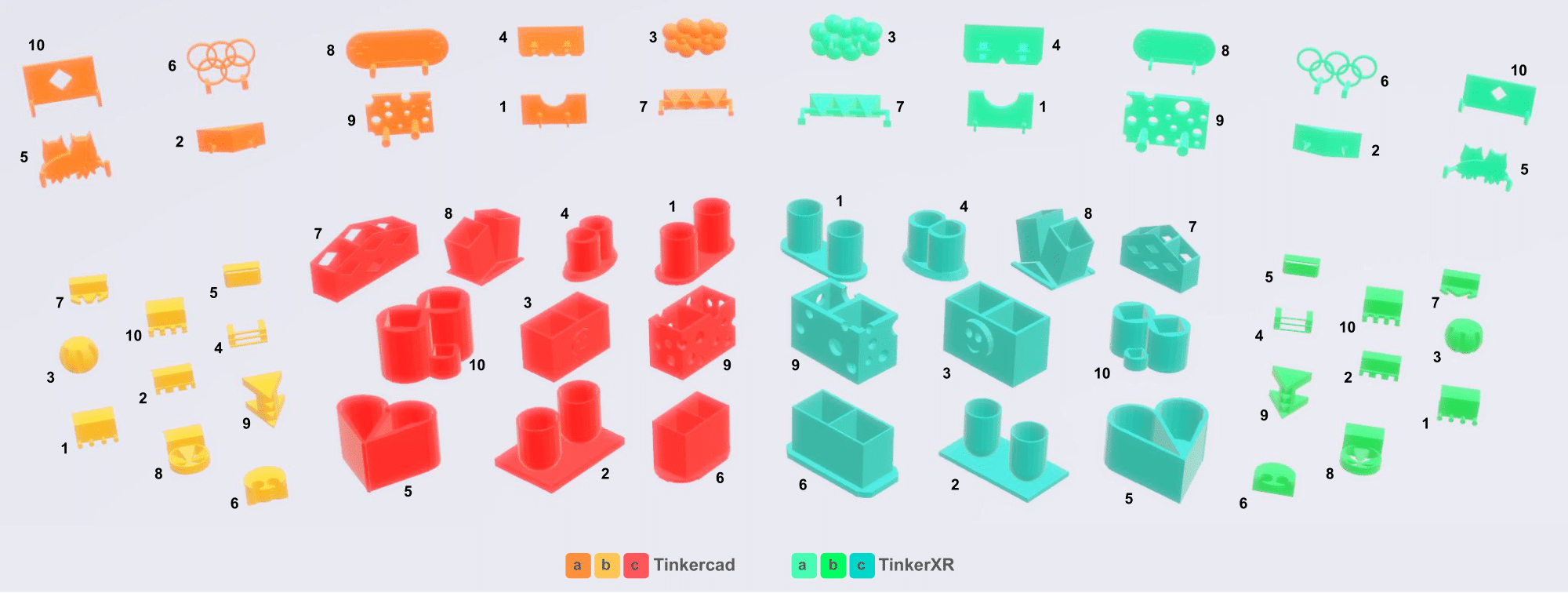}
  \caption{Designs created by the participants in the user study. Objects are colored based on their task number (a: wall-mounted key hanger, b: cable management, c: toiletries holder) and the system used for designing the models.}
  \Description{}
  \label{fig:user_study_objects}
\end{figure*}

\paragraph{Task \& Procedure}
For the evaluation, we conducted a within-subjects comparative study to assess \systemName against \textit{Tinkercad}. Each participant completed modeling tasks using both tools, with the order of tool usage counterbalanced across participants to mitigate order effects (e.g., half of the participants used \systemName first, followed by \textit{Tinkercad}, while the other half did the opposite). 
We showed both systems briefly, and participants completed a 3-minute warm-up before starting the tasks to familiarize themselves with the tools.
Participants were assigned three modeling tasks, each tailored to a specific context and accompanied by explicit design requirements:    
\textit{\#1 Wall-mounted key hanger}: Participants created a key hanger with two hooks narrow enough to accommodate keychain rings. This  task emphasized designing small, precise features and required the use of a vertical wall as the design context. 
\textit{\#2 Cable management}: Participants designed a holder to organize cables on a desk. The design had to securely fit the provided cables, ensuring they would not fall, and the organizer needed to remain stable when attached to the desk.
\textit{\#3 Toiletries holder}: Participants designed a holder for a toothpaste tube and toothbrush, intended for use in a bathroom. They were provided with physical objects representing the toiletries and used the \textit{reference object} functionality to ensure the holder dimensions aligned with the objects. For both conditions, participants were allowed to use a physical ruler to measure physical dimensions, although they were also allowed to use the digital AR ruler in the \systemName condition. 

Participants engaged with \systemName using a \textit{Quest 3} headset and hand interactions. For \systemName, they were instructed to prepare and start their prints on a virtual twin of the 3D printer; while for the \textit{Tinkercad} condition, they used \textit{Cura}\footnote{https://ultimaker.com/software/ultimaker-cura/}'s desktop UI for 3D printing preparation. To save time, participants did not wait for the physical prints to finish before moving on to the next task. The 3D printer \textit{slicing} sequence was only carried out for the first task due to the same reason. Each modeling session was followed by a survey to gather qualitative and quantitative feedback on the tools, focusing on usability, efficiency, and overall satisfaction. 

\paragraph{Measures}
We used both quantitative and qualitative metrics to evaluate the system.
We recorded the time required to complete the individual 3D \textit{design} tasks as a measure of overall system performance. For Task 1, we also measured the time required to transfer the finished design into the 3D \textit{printing} environment and prepare it for fabrication, i.e., \textit{slicing}. We measured this duration until the participant hit the print button.
After completing the tasks, participants were asked to complete standardized questionnaires, including the \textit{System Usability Scale} (SUS) \cite{brooke_sus_1996}, which we extended with additional questions related to the immersive experience,
and the \textit{NASA Task Load Index} (NASA-TLX) \cite{hart_development_1988}.
We collected qualitative feedback through open-ended questions regarding their overall experience.

\subsection{Results}

\subsubsection{Observations and Analysis of the Fabricated Artifacts}
\autoref{fig:user_study_objects} shows the three types of objects designed and fabricated by the participants. Each design task had specific requirements to ensure functionality and usability, and all participants successfully met these requirements. For the \textit{key hanger}, the design required at least two hooks to hold the keychains. The hooks had to be narrow enough to accommodate standard keychain rings. For \textit{cable management}, participants needed to design a holder that would securely fit at least two cables of the diameters provided, to ensure that the cables would not slip out. The design also had to accommodate the thickness of the desk to remain stable when attached. For the \textit{toiletries holder}, the design required at least two compartments or holes to hold the provided toothbrush and toothpaste tube. These compartments had to match the given dimensions to ensure a stable hold.
All models were validated by checking if they adhered to the defined dimensional requirements and functionality constraints. %  for each task. 

A notable insight was that cube primitive as a solid and cylinder geometry as a hole were mostly used objects for both the \systemName and \textit{Tinkercad} conditions.
During  \systemName usage, \textit{snap-to-grid} was heavily utilized. All participants used the physical ruler for the \textit{wall-mounted key hanger} and \textit{cable management tasks} while designing with \textit{Tinkercad}, while all used the reference object functionality for the \textit{toiletries holder} while designing with \systemName. Participants relied heavily on \textit{depth occlusion} and the highlighted projection of the objects on the grid to understand the location and depth of the objects.
Seen in \autoref{fig:user_study_objects}, the outcomes that have taken the most time on average of the two systems were the \textit{key hanger by P5} (18m 45s), the \textit{toiletiers holder by P5} (15m 31s), and the \textit{key hanger by P9} (14m 35s).
We note that, as seen in \autoref{fig:user_study_objects}, the majority of the resulting objects created with \systemName vs. \textit{Tinkercad} were comparable, i.e., none of the users created designs that differed deeply depending on the tool used. This verifies that  \systemName was able to offer similar 3D CAD capabilities despite the newly introduced immersive AR environment and not being an established commercial solution as a research artifact.

\begin{figure*}[t]
  \centering
  \includegraphics[width=0.7\textwidth]{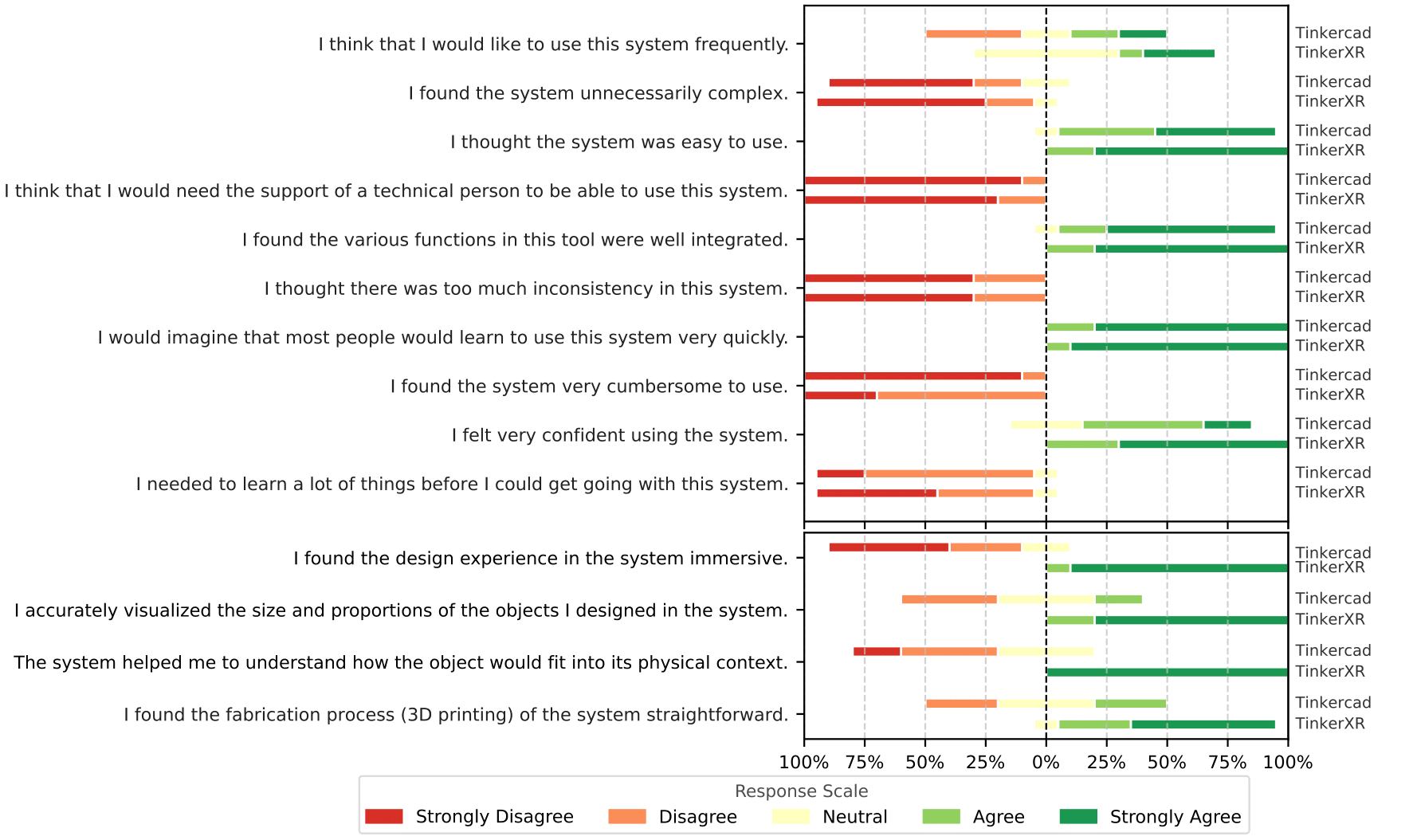}
  \caption{Results of the extended SUS survey comparing \textit{Tinkercad} and \systemName. User ratings are shown on a Likert scale from 1 (strongly disagree) to 5 (strongly agree).
  \systemName demonstrates generally higher ratings for immersive design features.}
  \Description{Results of the SUS survey.}
  \label{fig:sus_chart}
\end{figure*}

\subsubsection{Time}

Across all three \textit{design} tasks, we found no statistically significant differences in task completion time using \systemName compared to \textit{Tinkercad}: Task 1 (t = 0.272, p = 0.792), Task 2 (t = 0.272, p = 0.792), Task 3 (t = -1.769, p = 0.110). This suggests that \systemName provides an immersive design experience without a substantial time cost. However, future work may explore this tradeoff in larger or more varied user populations, e.g., for experts as well.
When it comes to the \textit{slicing} process for Task 1, the 3D printing preparation took less time when participants utilized \systemName. A paired t-test confirmed significantly shorter completion times in the \systemName condition for Task 1 (t = 3.747, p = 0.004).

\subsubsection{Surveys}

For the standard SUS survey, we obtained a score of 90\%, which shows that  \systemName system is within the \textit{A} grade (90\% or above), while \textit{Tinkercad} achieved a score of 85\%, which places it within the \textit{B} grade (from 80\% to 90\%) ~\cite{Sauro_Lewis_2016}.  
As shown in Fig. \ref{fig:sus_chart}, \systemName has higher scores across statements such as Q3, Q9, and Q10. 
These findings support that \systemName was easier to use for users in terms of coherence.
The additional questions related to immersive experience show that \systemName helped experience a more natural and hands-on design procedure.

Analysis of the NASA-TLX survey with a paired t-test reveals significant differences between \textit{Tinkercad} and \systemName across multiple workload dimensions. Participants reported lower \textit{mental demand} when using \systemName compared to \textit{Tinkercad} (t = 3.857, p = 0.004), indicating that the immersive AR environment facilitated easier comprehension and interaction during design tasks.
In contrast, \systemName exhibited higher \textit{physical demand} (t = -5.658, p \textless 0.001) and \textit{effort} (t = -2.684, p = 0.025), likely due to the dependence on hand gestures and the physical interaction required with the AR system compared to the mouse-and-keyboard interface of \textit{Tinkercad}. However, \textit{temporal demand}, \textit{performance}, and \textit{frustration} levels were comparable between the two, indicating that the usability benefits of \systemName may mitigate the added physical strain. 

Additionally, considering the current capabilities and limitations, users indicated that they would use \systemName for future projects, with an average rating of 4.4 on a 5-point Likert scale. They also recommended this tool to other novice users, giving it an average rating of 4.7 out of 5.

\begin{figure*}[t]
  \centering
  \includegraphics[width=0.5\textwidth]{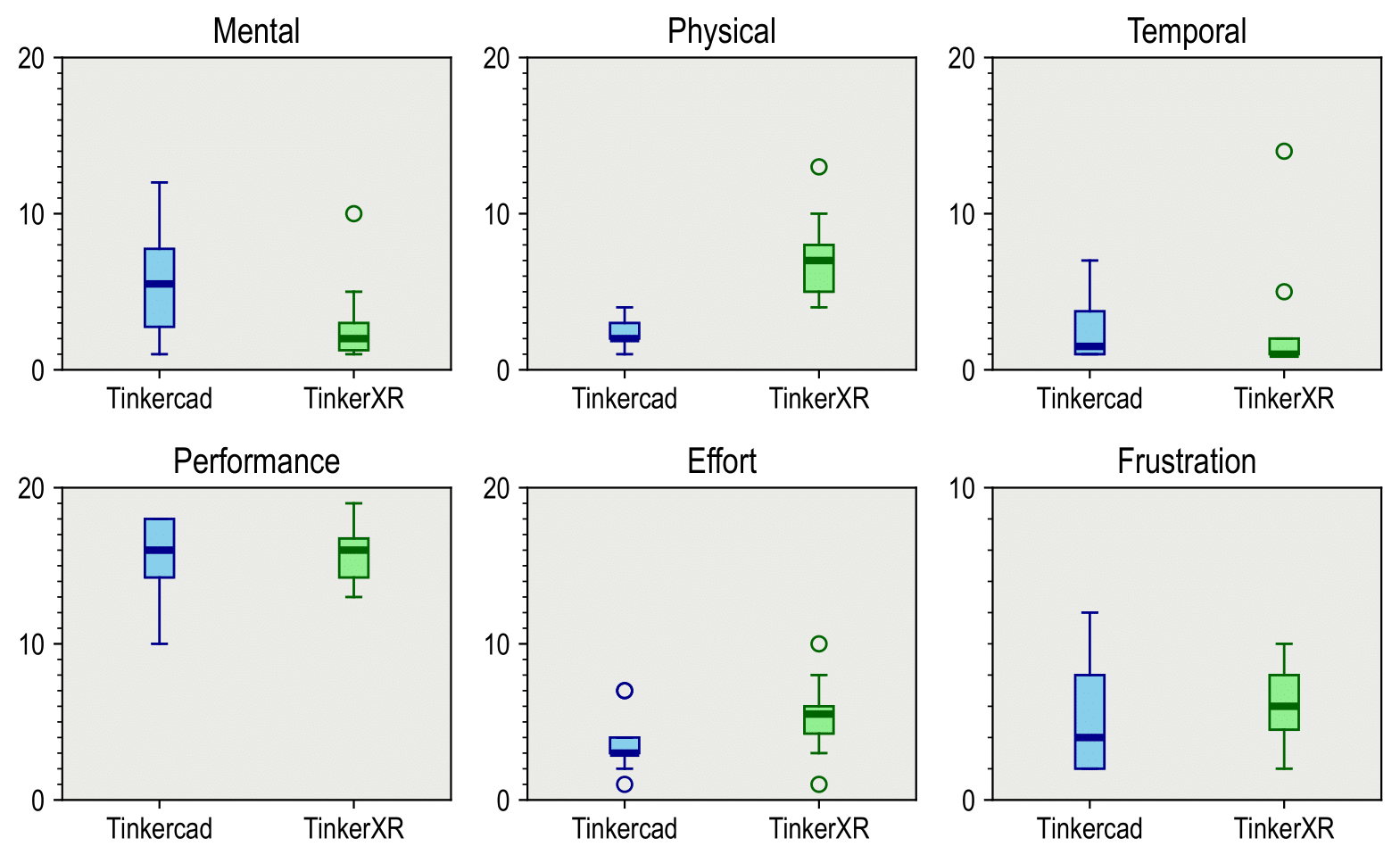}
  \caption{Results of the NASA-TLX survey comparing \textit{Tinkercad} and \systemName across six workload dimensions. \systemName shows lower mental demands and higher perceived performance, while \textit{Tinkercad} exhibits lower physical demand. Performance, temporal, and frustration scores are comparable between the two tools.
  }
  \Description{Result of the NASA-TLX survey.}
  \label{fig:nasa_chart}
\end{figure*}

\subsection{Qualitative Feedback}

Open-ended questions in our survey, such as \textit{"What did you like most about using the tool?"}, \textit{"What challenges did you face while using the tool?"}, and \textit{"What aspects of the tool would you most like to see improved?"},
and the thematic analysis of the responses allowed us to gather deeper insights into the system's strengths and areas for improvement under five overarching themes:

\paragraph{Real-World Integration and Immersive Experience}
Participants consistently praised the ability to design within real environments, emphasizing the value of \systemName's AR capabilities in visualizing objects in situ.
For instance, P1 shared, “\textit{With \systemName, I could see how my design would actually look in the space it was meant for}.”
P4 highlighted how designing anywhere — including outdoors — opened new possibilities, such as creating projects in natural settings. The immersive experience was widely appreciated, with features like depth occlusion enhancing realism. As P10 noted, “\textit{The occlusion made everything feel so much more immersive, like the objects were truly part of the room}.” The engaging and enjoyable nature of the system was also mentioned, with P3 stating, “\textit{Using the system was genuinely fun and engaging. I felt immersed}.”

\paragraph{Natural Controls}
The hand-based interactions stood out as significant strengths. P6 described them as “human-nature,” explaining, “\textit{Locking axes was so simple, just a (weak-hand) finger pinch and moving your hand on a direction}.” Similarly, the snap-to-grid feature was highly valued for its precision, with P2 commenting, “\textit{snap-to-grid was a game changer for precise alignments}.” and P3 commenting, "\textit{Snapping was very nice since it helped against my hand shaking on air}." These mechanisms allowed participants to focus on creativity without being hindered by complex control schemes.

\paragraph{Physical Demands and Ergonomics}
While the controls felt natural, several participants expressed concerns about the physical demands of prolonged use. P1 mentioned, “\textit{The physical demand was rather tiring},” and P2 noted, “\textit{It was a little tiring to have my arm raised for every action}.” These observations suggest need of improvements on reducing the physical effort required to perform operations in \systemName. It can be argued that with the current system, extended use could lead to fatigue, especially when designing complex models.

\paragraph{Accessibility and Future Adoption}
\systemName was widely commended for its ease of use and intuitiveness, even among novice users. P6 shared, “\textit{As someone completely new to AR, 3D modeling, and 3D printing, I was surprised by how easy it was to learn}.” This ease of use motivated participants to envision future applications, such as designing custom objects for their homes or gardens. P4 expressed excitement about using the system for personal projects, stating, “\textit{Even in my garden, I am already thinking about designing a birdhouse and some tree decorations}.” Most participants indicated they would continue using the tool and recommend it to others, with P8 emphasizing a room for a collaborative design directon: “\textit{This was a fun experience overall. I would recommend people to try it out. I wish I could design simultaneously with others in the same room too -- that would have been awesome}.”

\paragraph{Feature Expansion and Precision Tools}
On top of collaboration features, participants identified areas where additional design features could enhance the workflow. Several requested better tools for manipulating small objects, with P7 suggesting a “\textit{magnifying glass tool for designing small objects} and P1 mentioning "\textit{Small objects was hard to select and manipulate sometimes because I needed to be very precise with my hand}."  Building on the AR ruler feature, on-screen measurements were another commonly desired functionality, as P1 noted, “\textit{Length values of the edges appearing next to the objects, just like in Tinkercad, would help}.” Other recommendations included utilizing \textbf{generative design} and \textbf{speech recognition} to further simplify the design process. P4 also proposed more direct interaction methods, such as “\textit{grabbing objects directly with our hands (palm grab)}.”

\vspace{0.1cm}
Overall, the feedback highlights \systemName's strengths in providing an immersive and accessible design experience, especially for novices. However, challenges like solving the problem of high physical demand should be solved in the future. Addressing the lack of precision while manipulating small objects would further enhance the system's usability.

\section{Discussion}
\label{discussion}
In this section, we evaluate the effectiveness and future potential of \systemName. We discuss its suitability for both novice and expert users and express limitations related to handling complex 3D geometries. We also address the impact of the AR form factor on user experience and the limitations of hand gesture recognition. Additionally, we anticipate future advancements that could improve the system's capabilities and user interaction.

\paragraph{Novice vs. Expert Usage}
Our tool was designed with novices in mind, with the aim of providing an intuitive interface that simplifies both the design and fabrication of models. However, we envision that \systemName can also be beneficial to experts. Practitioners and professional designers can use it for rapid prototyping, concept development, and interactive presentations. The ease of use and flexibility of our system can streamline workflows, allowing experts to focus on more complex design tasks without being bogged down by technical logistics. In the future, we believe that our system can be adopted by educational institutions for teaching design principles, by industries for quick mock-ups or collaborative sessions, and by hobbyists seeking to explore 3D modeling and fabrication.

\paragraph{3D Geometry Complexity}
In Section~\ref{sec:applications}, we demonstrated our tool's usefulness through various design examples, each showing a different use case. These objects featured distinct geometries, ranging from small details to larger-scale structures. For instance, the candy bowl model in \autoref{fig:applications}d is made of 32 curved lines, each made of 8 cuboids, adding up to a total of 256 basic building blocks (if none of the blocks are combined) and 2048 vertices. Our system supports up to 300 basic building blocks and a total number of vertices of 20,000 in real time. However, we observed a noticeable lag when exceeding these numbers, especially in \textit{combination} operations. We anticipate this limitation can be addressed with future headsets offering higher processing power. Alternatively, more powerful headsets, e.g., \textit{Varjo XR-4}\footnote{\url{https://varjo.com/products/xr-4/}, Accessed: 4.08.2024}, which is connected to a desktop computer, can support compute-intensive tasks.

While the system can create a variety of complex geometries, certain tasks remain challenging due to the lack of more advanced tools such as extrusion of 3D geometries from 2D blueprints (e.g.,~sketched shapes, text), parametric curves, lofting, filleting, freeform design tools similar to those in \textit{Autodesk 3ds Max}\footnote{\url{https://www.autodesk.com/products/3ds-max}, Accessed: 4.08.2024}, or a collaborative structure allowing multiple users to work on the same project similar to \textit{Gravity Sketch} or adding physical engagement with local users and their surroundings similar to \textit{HoloBots}~\cite{ihara_holobots_2023}. While it is arguable that adding such features may increase complexity, their addition would elaborate the design capabilities, and should be considered as a future improvement.

We only implemented the features that were required for realizing the artifacts illustrated in Section~\ref{sec:applications} and objects with similar structural complexity, which were inspired by \textit{Tinkercad} repositories and related work. In the future, because we open-source our work, we believe that the community can add more features to enhance its capabilities. We anticipate future research to also support more advanced workflows, such as multi-material or -texture prints~\cite{ozdemir_speed-modulated_2024}. By collaborating with the community, we hope to see our system evolve to meet the needs of users of all proficiency levels.
%, ultimately broadening its applicability and usability.

\paragraph{Scaling to Other Fabrication Methods}
We believe that \systemName has the potential to adapt to other digital fabrication methods such as molding~\cite{serrano_mold-it_2022, valkeneers_stackmold_2019}, CNC machining~\cite{tran_oleary_imprimer_2023, tran_oleary_machine-o-matic_2019}, or laser cutting~\cite{dogan_structcode_2023, dogan_sensicut_2021} in the future. However, we acknowledge that this adaptation goes beyond simple file format conversion. Each method introduces unique design constraints, such as draft angles for molds, toolpath limitations for CNC milling, and material thickness or 2D geometry constraints for laser cutting. These factors would require the system to integrate CAM tools to guide users in adhering to method-specific constraints.

In this context, \textit{kyub}~\cite{baudisch_kyub_2019} provides inspiration for adapting 3D design for laser cutting. Exploring ways to integrate such techniques into \systemName could enable users to understand fabrication-specific constraints in real time, which we consider an exciting avenue for future work.

\paragraph{AR Form Factor}
MR headsets can be bulky; however, in our user study, participants did not report major problems due to the form factor. We used the  \textit{Quest 3} because it is lightweight, making it easy to use. In the future, we imagine AR technology becoming more ubiquitous while offering a more streamlined experience with the advent of AR glasses, such as \textit{Meta Orion}\footnote{\url{https://about.meta.com/realitylabs/orion/}, Accessed: 20.01.2025}. Future devices are likely to be more compact, resembling ordinary eyewear, thus enhancing user comfort and accessibility~\cite{dogan_ubiquitous_2024}. These advancements could lead to broader adoption and more immersive systems~\cite{imprinto}. Currently, \systemName is implemented exclusively for the \textit{Meta Quest} ecosystem. However, its core architecture, built in Unity, is fundamentally portable. Expanding the system to other platforms with robust hand tracking and spatial mapping capabilities, such as the \textit{Apple Vision Pro} or \textit{Varjo} headsets, represents a clear path for future development and would broaden the system's accessibility.

\paragraph{Hand Gestures}
While we are constrained by \textit{Meta}'s hand gesture APIs, which are known to improve over time\footnote{\url{https://www.techradar.com/computing/virtual-reality-augmented-reality/your-meta-quest-3-is-getting-a-hand-tracking-upgrade-that-could-unlock-foot-tracking}, Accessed: 5.08.2024}, there are still issues such as improper functionality in dimly lit areas or minor precision errors, especially during fine-scale manipulation. Our study showed that participants effectively used the snap-to-grid feature to mitigate this. To build on these strategies, future work could explore several paradigms to enhance precision further. For example, implementing a \textit{virtual magnifying glass} tool, as suggested by user feedback, would offer a stable, magnified view for detailed edits. Similarly, enabling users to fluidly work at a larger, more manageable scale before uniformly shrinking their design could simplify the manipulation of intricate components. Finally, extending the system to support dynamic \textit{object-to-object snapping} (e.g., aligning vertices, edges, and faces) would bring its precision capabilities closer to traditional desktop controls.

\paragraph{Utilization of Generative AI}
Our findings underscore the effectiveness of CSG for novices, yet the \textit{3D Geometry Complexity} discussion highlights its inherent limitations in creating freeform or highly intricate models. A compelling future research direction is the integration of generative AI to address this expressivity gap. In such a hybrid workflow, a user could leverage \systemName's direct manipulation capabilities to sculpt a low-fidelity "scaffold" of their design. 
This user-authored geometry, serving as a strong structural prior, could then guide state-of-the-art generative models % (e.g., Microsoft's \textit{Trellis}\footnote{\url{https://microsoft.github.io/TRELLIS/}, Accessed: 8.09.2025}, Tencent's \textit{Hunyuan}\footnote{\url{https://hunyuan-3d.com/}, Accessed: 8.09.2025}) 
to refine and detail the final object. Recent developer access to \textit{Quest~3}'s cameras\footnote{\url{https://github.com/oculus-samples/Unity-PassthroughCameraApiSamples}, Accessed: 8.09.2025} also opens the opportunity for these models to use environmental imagery to ensure the final design is aesthetically consistent with the user's room. This approach would preserve user \textit{agency} where the foundational design intent remains user-driven and AI acts as a tool for complex refinement rather than the sole creator~\cite{improvmate}. Investigating this interplay between direct manipulation and AI-driven detailing, and its effect on novice users' sense of authorship presents an interesting research question for the next generation of XR design tools.

\section{Conclusion}

In this paper, we presented \systemName, a new 3D modeling and fabrication tool through AR, designed for novices. Our approach lets users design artifacts with a simple primitive geometry manipulation CSG design logic, with intuitive hand gesture control schemes, on the intended space of the artifact, and with seamless integration with the 3D printing process. We showed the various features present in \systemName from selecting the workspace to design and fabrication. Using these features, we presented a wide variety of creations for different scenarios, each varying in complexity and practical applications.
A user study revealed that the system was effective for novices, and more immersive than \textit{Tinkercad}, while offering a comparable level of functionality and simplicity.
We hope that our open-source project will stimulate further exploration within communities engaged in HCI, design, and fabrication.

\begin{acks}
This work was supported by the \textit{INVERSE} Project under Grant No. 101136067, funded by the European Union.
\end{acks}

%%
%% The next two lines define the bibliography style to be used, and
%% the bibliography file.
\bibliographystyle{ACM-Reference-Format}
\bibliography{0-main}

\end{document}